%% file: alcalaAA-2003-0099.tex
\documentclass{aa}

\usepackage{graphicx}
\usepackage{lscape}
\usepackage{amssymb}

\begin{document}

\title{The Capodimonte Deep Field\thanks {Based on observations carried out 
at the European Southern Observatory, La Silla, Chile under proposals 
numbers 63.O-0464(A), 64.O-0304(A), 65.O-0298(A) and 67.B-0457(A).} }

\subtitle{Presentation of the survey and first follow-up studies}

\author{
 J.M. Alcal\'a\inst{1} 
 \and M.~Pannella\inst{2,1}
 \and E.~Puddu\inst{1}
 \and M.~Radovich\inst{1}
 \and R.~Silvotti\inst{1}
 \and M.~Arnaboldi\inst{3}
 \and M.~Capaccioli\inst{1,4}
 \and G.~Covone\inst{8,1}
 \and M.~Dall'Ora\inst{7,1}
 \and G.~De Lucia\inst{5,1}
 \and A.~Grado\inst{1}
 \and G.~Longo\inst{4,1,10}
 \and A.~Mercurio\inst{1}
 \and I.~Musella\inst{1}
 \and N.~Napolitano\inst{6,1}
 \and M.~Pavlov\inst{1}
 \and A.~Rifatto\inst{1}
 \and V.~Ripepi\inst{1}
 \and R.~Scaramella\inst{9}
}

\offprints{J.M. Alcal\'a (jmae@na.astro.it) }

\institute{INAF-Osservatorio Astronomico di Capodimonte, Via Moiariello 16,
           I-80131, Napoli, Italy
\and
          Max-Planck Institut f\"ur Extraterrestrische Physik, 
          Giessenbach str., Garching bei M\"unchen.
\and
          INAF-Osservatorio Astronomico di Torino, Via Osservatorio 20
          I-10025 Pino Torinese, Italy
\and
          Dipartimento di Scienze Fisiche, Universit\`a Federico II, Napoli, Italy 
\and      
          Max Planck Institut f\"ur Astrophysik, D-85748 Garching, 
          Germany
\and 
          Kapteyn  Astronomical Institute, Landleven 12, 9747 AD Groningen, 
          The Netherlands
\and
          Dipartimento di Fisica, Universita' di Roma Tor Vergata, Via della
          Ricerca Scientifica 1, 00133 Rome, Italy.
 \and
          Laboratoire d'Astrophysique de Marseille B.P. 8  - 13376 Marseille 
          Cedex 12, France
 \and
          INAF-Osservatorio Astronomico di Monteporzio, Via di Frascati 33, 
          I-00044, Roma, Italy
 \and     
           INFN - Napoli Unit, via Cinthia 6, 80126 Napoli, Italy 
}

\date{Received: ... ; Accepted: ...}

\abstract{ 
 We present the Capodimonte Deep Field (OACDF), a deep field covering an 
 area of 0.5 $\deg ^2$ in the $B$, $V$, $R$ optical bands plus six medium-band 
 filters in the wavelength range 773--913 nm. The field reaches the following 
 limiting magnitudes: $B_{AB}\sim 25.3$, $V_{AB}\sim 24.8$ and $R_{AB}\sim25.1$
 and contains $\sim$ 50000 extended sources in the magnitude range 
 18 $\le R_{AB} \le$ 25.0. Hence, it is intermediate between deep 
 {\it pencil beam} surveys and very wide but shallow surveys. 
 The main scientific goal of the OACDF is the identification and 
 characterization of early-type field galaxies at different look-back 
 times in order to study different scenarios of galaxy 
 formation. Parallel goals include the search for groups and clusters of 
 galaxies and the search for rare and peculiar objects (gravitational lenses, 
 QSOs, halo White Dwarfs). In this paper we describe the OACDF data 
 reduction, the methods adopted for the extraction of the photometric catalogs,
 the photometric calibration and the quality assessment of the catalogs by means 
 of galaxy number counts, spectroscopic and photometric redshifts and star 
 colors.
 We also present the first results of the search for galaxy overdensities. 
 The depth of the OACDF and its relatively large spatial coverage with respect
 to pencil beam surveys make it a good tool for further studies of galaxy 
 formation and evolution in the redshift range 0--1, as well as for stellar
 studies.
}

\maketitle

\section{Introduction}

Wide and deep surveys are a vital tool in modern astronomy: they provide 
the large datasets of objects needed (both at high and low redshifts)
to improve our knowledge of the physical mechanisms driving galaxy 
formation and evolution.

Recent wide field surveys, such as DPOSS, 2MASS, SDSS and 2DF,
(Djorgovski et al. 2000, Colless et al. 2001; Jarrett et al. 2000; 
York et al. 2000) provide an unprecedented wealth of information 
on the structure of the {\em local} Universe.

On the other hand, the complementary {\it pencil beam} deep surveys (i.e. 
relatively small patches of the sky observed in several bands with high 
angular resolution and high S/N ratios at faint magnitude levels) allow 
us to investigate the high-$z$ Universe ($z \gtrsim 1$), providing large 
sets of distant galaxies to constrain the different evolution scenarios.

In the past, the significance of deep fields was limited by the constraints
of the small field of view covered by single CCD detectors and the need 
to obtain large fractions of observing time with 4~m class telescopes. 
The small spatial coverage of the first studies 
(Tyson 1988, Lilly et al. 1995; Ellis et al. 1996; Le Fevre et al. 1996) 
succeeded in tracing only a rough and broad picture of 
galaxy evolution in the redshift range $0 < z < 0.8 $ and therefore did not 
allow strong constraints on the various scenarios proposed for galaxy 
formation and evolution (White and Frenk 1991; Cole et al. 2000; 
Aragon-Salamanca et al. 1998; Peebles 2002). 

The Hubble Deep Fields (e.g. Ferguson, Dickinson \& Williams
2000) has marked a cornerstone in our understanding of  
galaxy evolutionary processes, providing unprecedented contraints on the 
theoretical scenarios for galaxy formation and evolution.

Similar deep surveys have been recently performed using ground-based 
telescopes of the 10m class generation: the Subaru Deep Field 
(Maihara et al. 2001), the Fors Deep Survey (Heidt et al. 2003) and the 
Gemini Deep Deep Survey (Abraham et al. 2004).
However, an important drawback of these deep surveys is that the 
typical field of view has angular sizes at $z \sim 1 $ which are relatively 
small with respect to the scales relevant for large-scale structure.
Thus, in order to study the properties of distant galaxies at high redshift 
in different large scale enviroments, it is mandatory to control 
field-to-field variations, by means of deep surveys along many different 
lines of sight or extended to larger contiguous sky areas.

The advent of wide field imagers on dedicated survey telescopes, like the 
wide-field imager (WFI) at the ESO/MPG 2.2m telescope, allow one to perform deep 
surveys on sky areas of the order of 1 square degree, thus providing an ideal 
tool to gather high accuracy data for statistically significant samples of 
objects in the redshift range $0<z<1$.
This has prompted the observation of several deep fields at optical and near 
infrared wavelengths:  COMBO-17 (Wolf et al. 2003); K20 (Cimatti et al. 2002), 
MUNICS (Drory et al. 2001), the VIRMOS-VLT deep field (Le Fevre et al. 2003; 
Mc Cracken et al. 2003; Radovich et al. 2004).

While these surveys cannot, in general, compete with the HDFS and the 10m 
class telescope deep surveys in terms of depth, they allow one to study in 
detail the galaxy population up to $z \simeq 1$ in a much larger field of view.

This fact is extremely important, for instance, when determining the 
luminosity function and its possible evolution up to $z \simeq 1$
(e.g., Wolf et al. 2003).

In 1999, the Osservatorio Astronomico di Capodimonte (INAF-OAC) started 
the Capodimonte Deep Field (OACDF) project using WFI at the ESO/MPG 2.2m
telescope at La Silla, Chile. The OACDF covers an area of 0.5 $\deg ^2$ 
in the $B$,$V$, $R$ optical bands and in six medium-band filters in the 
wavelength range 773--913~nm. The OACDF is intermediate between deep 
pencil beam surveys and very wide, but shallow surveys. As it will be shown 
in the following paragraphs the OACDF reaches limiting magnitudes 
$B_{AB}\sim 25.3$, $V_{AB}\sim 24.8$ and $R_{AB}\sim25.1$ and contains 
$\approx$ 50000 extended sources in the magnitude range 
18 $\le R_{AB} \le$ 25.0. 

The OACDF was complemented with a first spectroscopic follow-up and with 
a shallow photometric survey overlapping the deep survey and covering 
1 $\deg ^2$ in the $B$,$V$, $R$ and $I$ bands. The shallow survey reaches 
a depth of $I_{AB}\sim22.1$ and was planned in order to enhance the 
statistics at low redshift. 

The OACDF observing strategy was driven mainly by the identification and
characterization of a statistically significant sample of early-type galaxies 
at different redshifts and in different environments. The number density, 
as well as the global morphological and photometric properties of these 
objects, are predicted to vary for different formation scenarios 
(Aragon Salamanca et al. 1998, Kauffmann \& Charlot 1998, 
Somerville et al. 2004). The filter set-up and in particular the availabilty 
of medium band filters in the range 753-913 nm, allow us to trace the D4000 
break over the redshift range from 0 to 1 and therefore to obtain an accurate 
estimate of the photometric redshifts for these objects.\\

Parallel goals of the OACDF project are:

\begin{itemize}

\item[-] 
 the search for groups and clusters of galaxies to study 
 the luminosity function of galaxy clusters, as well as their luminosity 
 and color segregation, both linked to the more general question of the 
 evolution of galaxies in environments with different densities 
 (Menanteau et al. 1999). Another aspect is to study the cluster morphology 
 (clustering and sub-clustering) which bear relevant information on the 
 history of the Universe (Navarro, Frenk and White 1996). The identification 
 of a sample of groups of galaxies and intermediate-richness clusters of 
 galaxies at intermediate redshift, extracted in a homogeneous way, is 
 necessary to study the properties of galaxies in moderately 
 dense environments. 
 
\item[-]
the search for rare/peculiar objects, in particular gravitational 
lenses and high-redshift QSOs (z$>$3); 

\item[-]
the identification of nearby galaxies with strong emission lines 
using the medium-band filters to perform clustering and 
correlation function studies.

\item[-]
to provide a photometric and astrometric database for stellar studies.

\end{itemize}

While some peculiar objects present in the OACDF have already been discussed 
elsewhere (cf. gravitational lenses in Sazhin et al. 2003; white dwarfs 
in Silvotti et al. 2003), in this paper we present the OACDF photometric 
calibration, the catalog extraction, the quality assessment of the photometric
catalogs through the derivation of galaxy number counts, spectroscopy of 
stars and galaxies by means of photometric and spectroscopic redshifts. 
We also present the first results of the search for galaxy overdensities. \\

In a forthcoming paper, we will use the OACDF data, the photometric 
redshift determinations and the available spectra in order to analyze the 
redshift and environmental evolution of scaling-laws, such as the Kormendy 
relation (Kormendy 1977, Ziegler et al. 1999, La Barbera et al. 2003), and 
of the stellar populations of these objects through the use of the Lick 
indices (Faber et al. 1985, Worthey et al. 1992) and stellar population 
synthesis models (Thomas et al. 2003).   

Other future OACDF papers will include the following tasks:

\begin{itemize}

\item[-]
search for groups and clusters of galaxies: in the present paper we describe
the adopted methods for the detection of groups and clusters of galaxies in
the OACDF (see Section~6). In a future paper we will use the candidate clusters
as part of a larger statistical sample needed to derive the multiplicity function
and to study the luminosity function as a function of the environment. 
Follow-up spectroscopic studies are in course.\\

\item[-]
search for Gravitational lenses: an important by-product of a survey like 
the OACDF is the possible detection of strong gravitational lenses (GLs).
We have therefore started an extensive search for GLs in the OACDF2 and 
OACDF4 fields by spectroscopic observations at the ESO 3.6m telescope at 
La Silla, Chile. In a future paper we will present the methodology and 
the results of this search and, in particular, its implications for wider
surveys with VST, which will allow us to discover several tens of bright GLs 
in the Southern emisphere.\\

\item[-]
the spectroscopic confirmation of white dwarf (WD) candidates 
selected through their colors. Spectroscopic data were obtained
at the ESO NTT and 3.6m telescopes and are under reduction; some preliminary 
results are shown in Silvotti et al. (2003). The main goal of this project is 
to increase the WD statistics, in particular for the cooler, 
fainter and older objects, which  contain crucial information on the genesis 
of our Galaxy: age of the galactic disk (and halo) through the WD luminosity 
function, initial mass function, and stellar formation rate 
(see Fontaine et al. 2001 for an updated review). Of particular interest is 
the halo WD population as these rare and faint objects, whose 
statistics are almost totally unknown, might partially contribute to the 
galactic dark matter. This is suggested by the results of the MACHO+EROS 
microlensing experiments which show that microlensing events are mainly 
produced by halo objects with an average mass of $\sim$0.5 M$_{\odot}$ 
(Alcock et al.~2000).

\end{itemize}

The present paper is structured as follows: in Section~2 we present 
the field selection and the observational strategy for the imaging 
and initial spectroscopic follow-up. 
The data reduction and photometric calibration, as well as the systematic 
behavior of the zero points of the ESO-WFI CCDs, and the spectroscopic 
data reduction are reported in Section~3. In Section~4 we discuss 
the problems posed by catalog extraction, while the quality assessment 
of the OACDF catalogs are discussed in Section~5. First results on
the search for groups and clusters of galaxies are reported in Section~6 
and a summary is presented in Section~7. Finally, the Appendix reports a 
catalog containing magnitudes and spectroscopic redshifts for 114 galaxies 
observed in our first spectroscopic follow-up. The catalog also contains 
the magnitudes and spectral types for 59 stars in the OACDF. 

\section{Field selection and observations}

The central coordinates ($\alpha$(2000)$\approx$12:25:10, 
$\delta$(2000)$\approx$-12:48:31) of the 
OACDF were selected in order to identify a field matching the following 
criteria:

\begin{itemize}
\item[{\it i)}]~lack of sources brighter than $V=9$ mag  to
  minimize CCD saturation and ghost effects;
\item[{\it ii)}]~high galactic latitude to avoid high
  stellar crowding;
\item[{\it iii)}]~low interstellar (IS) extinction ($E(B-V) < $0.03~mag) 
  according to the Burstein and Heiles (1982) maps;
\item[{\it iv)}]~to be observable from both hemispheres to have 
  the possibility of performing follow-ups using telescopes at several sites.
\end{itemize}

The observations were carried out in three different periods
(18--22 April  1999, 7--12 March 2000 and 26--30 April 2000),
using the Wide-Field Imager (WFI) CCD mosaic camera (Baade et al.
1998) at the ESO/MPG 2.2m telescope at La Silla in Chile. This
camera consists of eight 2k$\times$4k CCDs forming a 8k$\times$8k
array with a scale of $0.238''$/pix.

In the shallow survey, four adjacent $30'\times30'$ fields
(hereafter OACDF-1 through 4) covering a $1^{\circ} \times
1^{\circ}$ field in the sky were observed in the $B$, $V$, $R$
and $I$ bands (see Table~\ref{tab:centers} for the coordinates of
the centers of the $4$ fields). The OACDF2 and OACDF4 were also
selected for the 0.5~deg$^2$ deep survey, performed in the $B$,
$V$, $R$ broad-bands and in the $\lambda$753, $\lambda$770,
$\lambda$791, $\lambda$815, $\lambda$837, $\lambda$884 and
$\lambda$914 medium-band filters.

For the shallow survey, each field was observed in $5$ ditherings,
while at least 8 ditherings were obtained for the deep survey, the
exact number depending on the band. The adopted dithering pattern
is that described in the ESO-WFI observers guide (Brewer and
Agustein 2000). A detailed summary of the observations is provided
in Table~\ref{tab:obs}.

\begin{table}
\begin{flushleft}
\caption[]{\label{tab:centers} Coordinates of the OACDF centers.}
\begin{tabular}{lccccccccccc}
\hline
Field      & $\alpha$(2000)    & $\delta$(2000)   \\
           &                   &               \\  \hline
OACDF1     &  12 \ 26 \ 20.4 & -12 \ 30 \ 20 \\
OACDF2     &  12 \ 24 \ 27.0 & -12 \ 30 \ 20 \\
OACDF3     &  12 \ 26 \ 20.4 & -13 \ 01 \ 20 \\
OACDF4     &  12 \ 24 \ 27.4 & -13 \ 01 \ 20 \\
               &                 &           \\ \hline
\end{tabular}
\end{flushleft}
\end{table}

\begin{table*}
\begin{flushleft}
\caption[]{\label{tab:obs} Summary of the OACDF observations.}
\begin{tabular}{lcccccccr}
\hline
Field          & obs. run         & Filter & \# diths &Total exp.& Seeing &  Airmass & Zero points & Col.Terms \\
               &                    &        &          &  time    & arcsec &          & $ZP^\dagger$& \\
\hline
OACDF2 deep    &       1$^{st}$   &   $B$  &    12    &  2.0h    & 1.24   &  1.287 & 24.77         & $+$0.24  \\
OACDF2 deep    &       1$^{st}$   &   $V$  &    10    &  1.7h    & 1.07   &  1.184 & 24.30         & $-$0.14  \\
OACDF2 deep    &       1$^{st}$   &   $R$  &    13    &  3.3h    & 1.11   &  1.482 & 24.63         & $-$0.03  \\

OACDF2 deep    & 1$^{st}$-2$^{nd}$&  753nm &   9+10   &  6.5h    & 0.88   &  1.512 & 21.95         &          \\
OACDF2 deep    & 1$^{st}$-2$^{nd}$&  770nm &   9+10   &  6.0h    & 0.86   &  1.544 & 21.88         &          \\
OACDF2 deep    & 1$^{st}$-2$^{nd}$&  790nm &   9+10   &  6.5h    & 0.99   &  1.084 & 21.80         &          \\
OACDF2 deep    & 1$^{st}$-2$^{nd}$&  815nm &    9+9   &  6.8h    & 0.79   &  1.044 & 21.59         &          \\
OACDF2 deep    & 1$^{st}$-2$^{nd}$&  837nm &    9+8   &  6.6h    & 0.95   &  1.468 & 21.54         &          \\
OACDF2 deep    & 1$^{st}$-2$^{nd}$&  914nm &  10+10   &  5.6h    & 0.90   &  1.046 & 21.15         &          \\

OACDF4 deep*   &      3$^{rd}$     &   $B$  &    8     &  2.0h    & 0.99   &  1.095 & 24.76         & $+$0.22  \\
OACDF4 deep    & 2$^{nd}$-3$^{rd}$&   $V$  &    7     &  1.8h    & 0.95   &  1.042 & 24.23         & $-$0.18  \\
OACDF4 deep    & 2$^{nd}$-3$^{rd}$&   $R$  &   4+14   &  4.2h    & 0.99   &  1.410 & 24.54         & $-$0.03  \\

OACDF4 deep    & 2$^{nd}$-3$^{rd}$&  753nm &   5+10   &  4.8h    & 0.95   &  1.042 & 21.81         &          \\
OACDF4 deep    &      3$^{rd}$     &  770nm &    7     &  2.1h    & 1.33   &  1.250 & 21.72         &          \\
OACDF4 deep    &      3$^{rd}$     &  790nm &    9     &  3.0h    & 0.95   &  1.380 & 21.65         &          \\
OACDF4 deep    & 2$^{nd}$-3$^{rd}$&  815nm &   5+6    &  3.5h    & 0.95   &  1.110 & 21.42         &          \\
OACDF4 deep    &      3$^{rd}$     &  837nm &    7     &  3.3h    & 1.20   &  1.100 & 21.44         &          \\
OACDF4 deep    &      3$^{rd}$     &  914nm &    9     &  2.5h    & 0.81   &  1.110 & 21.05         &          \\

OACDF shallow  &      1$^{st}$     &   $B$  &    5     & 20min    & 0.86   &  1.078 & 24.77         & $+$0.24  \\
OACDF shallow  &      1$^{st}$     &   $V$  &    5     & 10min    & 0.83   &  1.237 & 24.30         & $-$0.14  \\
OACDF shallow* &      2$^{nd}$     &   $R$  &    5     & 10min    & 1.38   &  1.320 & 24.45         & $-$0.03  \\
OACDF shallow  &      1$^{st}$     &   $I$  &    5     & 10min    & 0.81   &  1.650 & 23.31         & $+$0.12  \\

\hline
\end{tabular}
\end{flushleft}

$\dagger$ The reported zero points are those for the first and third observing runs,
as the second run was partially non-photometric.

* Due to technical problems it was not possible to calibrate the $B$-band
during the third run. Stars on the OACDF4, calibrated during the first
observing run, were used as secondary standards. A similar procedure
was applied to the shallow $R$-band images of the second run.

\end{table*}

The standard fields L\,98, L\,104, L\,107, L\,110 and PG\,1525, selected from 
the Landolt (1992) equatorial regions, were observed to perform 
the broad-band photometric calibration. 
Likewise, the spectrophotometric standard stars EG\,274, Hill\,600, LTT\,3864,
LTT\,4364,  LTT\,6248, and LTT\,7379 were observed in the medium-band filters 
for absolute flux calibration purposes. Typically, each standard field was 
observed twice.

\subsection{Spectroscopic observations}

The first spectroscopic follow-up of the OACDF was performed in April 2000 
using the ESO Multi-Mode Instrument (EMMI) at the New Technology Telescope 
(NTT) in the multi-object spectroscopy (MOS) mode. The selection criteria 
for the observed objects will be published in our forthcoming paper on the 
identification and characterization of early-type galaxies in the OACDF. 
In this paper we report the observations
and use the spectra to fine-tune the software parametrization for the 
photometric redshift determinations and for quality assessment.

Five 9 x 4 arc-min$^2$ fields distributed in the OACDF2 were observed through 
grism No. 3, which yields a dispersion of 2.3 \AA/pix. Some ten flat-field 
exposures were taken nightly. Before and after the science spectra, HeAr 
spectra were observed in each configuration and mask for wavelength 
calibration purposes. Given the faint magnitude of the targets, the pixels 
in the dispersion direction were re-binned by a factor of two 
to minimize the noise. The resulting nominal resolution was about 
10~\AA\ (FWHM) and the covered spectral range was from about 3800~\AA\ to 
about 8800~\AA, but the resulting spectral range for each individual object 
depends on the position of each individual slit relative to the edges of 
the mask.   
To remove cosmic ray hits and increase the S/N ratio, three 
science exposures of 40 minutes each were made, amounting to a total of 
2 hours of exposure per field. Three acquisition images, useful for the 
selection of the slit positions, were obtained by the NTT team prior to 
the observations. The other three acquisition images were obtained during 
the first observing night. 
Typically 35 slits per mask were used. Three apertures were normally 
assigned to bright stars for mask alignment purposes.
The spectrophotometric standard stars Hil600 and Eg274 were observed with 
the same instrumental set-up as the objects for flux calibration.
 
\section{Data reduction}

\subsection{Imaging}

Data reduction was performed using the task {\it mscred} in 
IRAF\footnote{IRAF is distributed by the National Optical Astronomy 
Observatories, which is operated by the association of the universities 
for research in astronomy, Inc., under contract with the National Science 
Foundation.}. Bias correction, super-flat-fielding and fringing correction 
were done on a nightly basis.

For each dithering sequence we chose a reference exposure and
computed the astrometric solution for the CCDs in that exposure
with reference to an external astrometric catalog. A polynomial
fit was then performed between the source positions in the
reference CCD and those in the other exposures. As reference
astrometric catalogue we use the USNO--A2 catalog (Monet et al.
1998) which allows to achieve an absolute RMS accuracy $\sim$
0.3". In order to improve the internal astrometric precision we 
first used the USNO--A2 catalog as a reference for the R-band and 
the resulting catalog, extracted from the R-band stacked image, 
was then used as reference catalog for the 
pointings in the other bands. The internal RMS accuracy between 
source positions in different bands is less than 0.1". In order 
to correct for transparency variations we also selected a reference 
image and normalized the others to the same flux level, using sources 
in the overlap regions. Finally, images were re-sampled, scaled in 
flux and co-added using the average sigma clip. More details in the 
WFI data reduction are provided in Alcal\'a et al. (2002).

\subsubsection{The photometric calibration}

The broad-band photometric calibration to the Johnson-Cousins
$BVR_CI_C$ system was performed using the standard Landolt (1992)
fields. For the field Landolt-98, we used the larger set of photometric 
standards by Stetson (2000). Photometric calibrated magnitudes and 
instrumental magnitudes are related by the following equations:
 
\begin{equation}
 B = b + C_B \cdot (B - V) +  Z_B,
\end{equation}

\begin{equation}
 V = v + C_V \cdot (B - V) +  Z_V,
\end{equation}

\begin{equation}
 R_{C} = r + C_R \cdot (V - R_{C}) +  Z_R,
\end{equation}

\begin{equation}
 I_{C} = ~i~ + C_I \cdot (V - I_{C}) + Z_I,
\end{equation}
 
\noindent 
 where $b, v, r, i$ are the instrumental magnitudes corrected 
 for atmospheric extinction, $C_B,C_V,C_R,C_I$ the color terms 
 and $Z_B,Z_V,Z_R,Z_I$ the zero points. We used the mean extinction 
 coefficients for the La Silla observatory
 \footnote{see http://obswww.unige.ch/photom/extlast.html and 
 http://www.ls.eso.org/lasilla/sciops/2p2/D1p5M/misc/ Extinction.html}.

\begin{figure*}  
\vspace{8.0cm}
\includegraphics{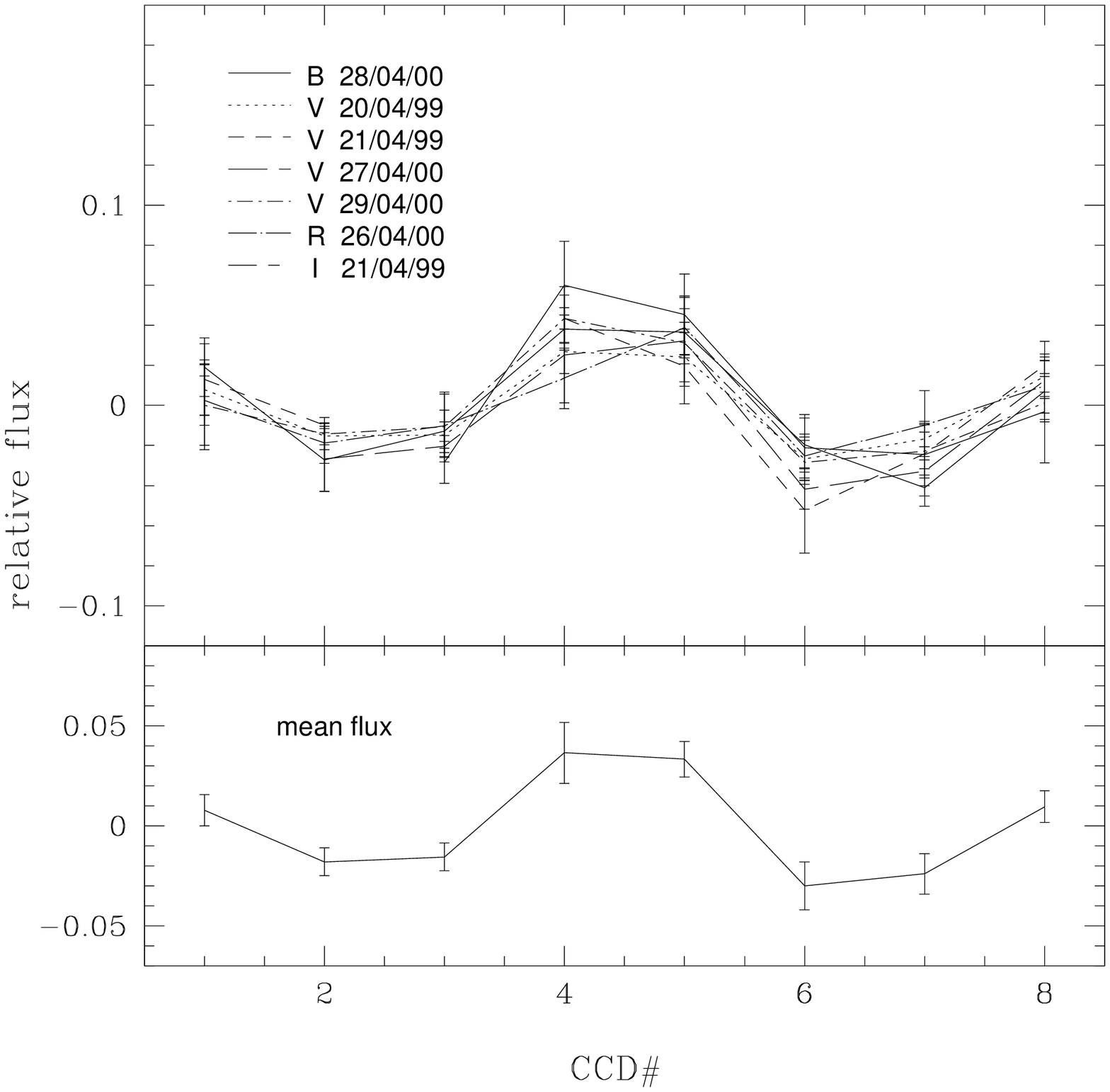}
\includegraphics{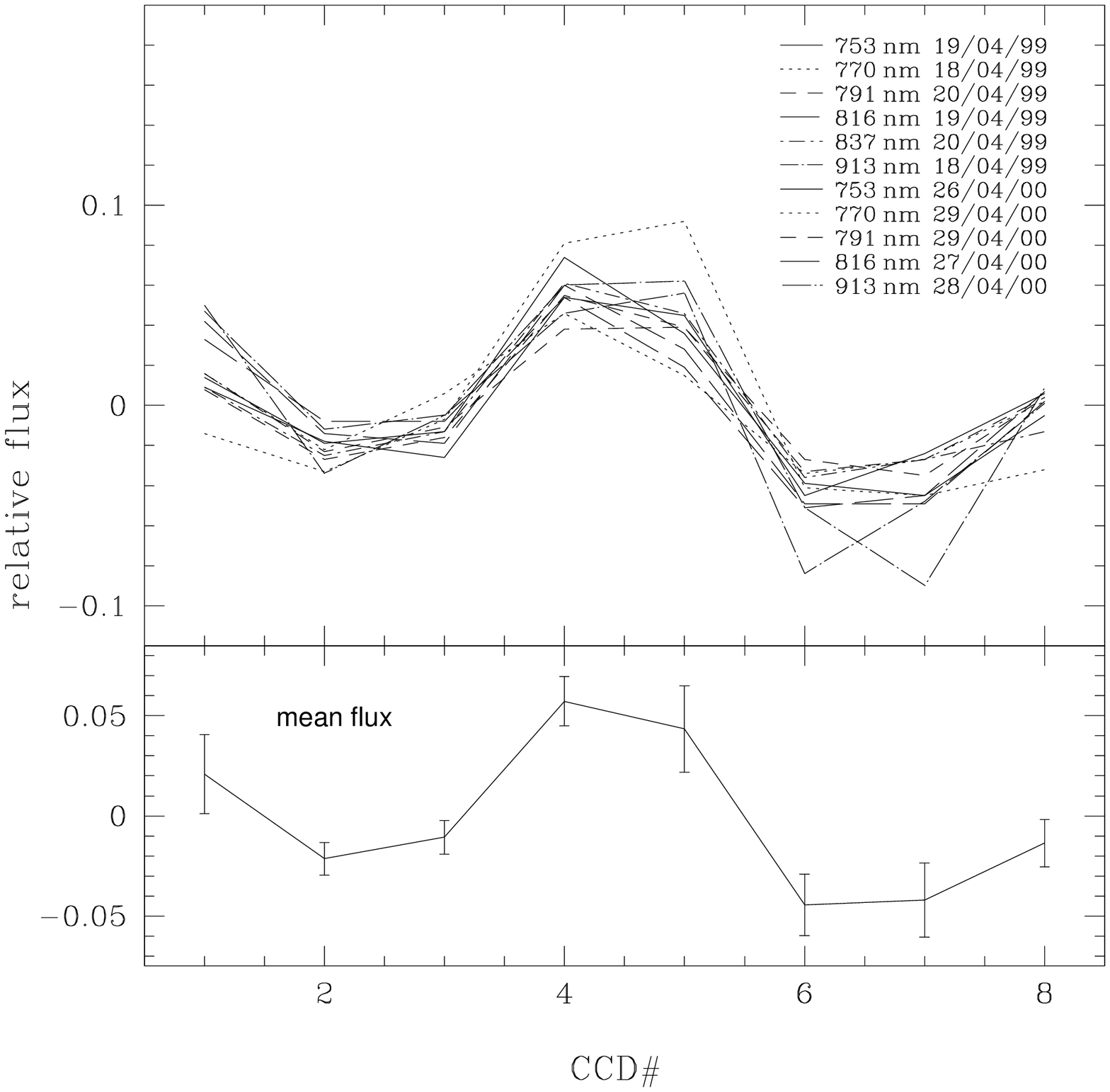}
\caption[]{\label{scatterlight} Left: relative photometric offsets 
versus the CCDs for the broad-band filters (upper panel). The different line 
types represent the different observing dates as labeled. The average over 
the broad-band filters is shown in the lower panel. Right: the same 
but for the medium-band filters.}
\end{figure*}

Photometric calibrations were achieved using a number of stars
ranging from $100$ to $200$. The fit residuals are typically less
than 3\%, but in some cases the residuals are as high as 5\%. The
relatively high residuals may be a consequence of the non-uniform
background illumination over the field of view, that will be
discussed in Section~3.2. A few points with particularly high
residuals are due to stars partially falling in the CCD gaps
and/or to photometric/astrometric errors in the standard catalog.
Aberrant points were rejected by a sigma clipping algorithm.
Points typically above the 2$\sigma$ level were not used for the
calibration.

The photometric zero points and the corresponding color terms are
listed in Table~\ref{tab:obs}.
Since the second run was partially non-photometric, we give in
Table~\ref{tab:obs} only the zero points of the first and third
observing runs.
Also, due to technical problems, it was not  possible to calibrate
the $B$-band of the field OACDF4 during the third observing run. The
latter calibration was done using secondary standard stars in the
OACDF.  A similar approach was adopted for the shallow observations
of the second observing run in the $R$-band.

The flux calibration for the medium-band observations was performed using 
spectrophotometric standard stars. These calibrations provide AB magnitudes. 
The detailed explanation of the medium-band filter calibrations, as well as 
the determination of AB magnitudes can be found in Alcal\'a et al. (2002). 

For practical purposes, we transform our broad-band instrumental WFI
magnitudes into AB magnitudes using the following expressions:\\

\begin{equation}
 B_{AB} = b +  Z_B - 0.093,
\end{equation}

\begin{equation}
 V_{AB} = v +  Z_V + 0.002,
\end{equation}

\begin{equation}
 R_{AB} = r +  Z_R + 0.215,
\end{equation}

\begin{equation}
 I_{AB} = ~i~ + Z_I + 0.509
\end{equation}

In the catalog presented in the appendix, AB magnitudes are given 
(Tables~\ref{tab:brmags} and \ref{tab:nrmags}).

In order to check the photometric homogeneity over the eight CCDs
and its dependence on the star color (Manfroid and Selman 2001;
Monelli et al. 2003), we also observed a few selected standard
fields in each of the eight CCDs. For the broad-band filters, we
used the Landolt field PG1525-071, which contains five stars with
$(B-V)$ colors ranging from -0.198 to 1.109. For the medium-band
filters we used the spectro-photometric standards Eg~274, Hill~600
and LTT~6248. We find no color dependence as well as no time dependence 
but, on the other hand, we find an offset of the instrumental magnitude 
among the eight CCDs. The effect is summarized in Figure~\ref{scatterlight}, 
where we show the fluxes of the standard stars measured in each CCD, 
relative to the mean value over the 8 CCDs. Within the errors, the trend 
is the same in the different bands ($B$, $V$, $R$, $I$) and at different 
times (from April 1999 to April 2000). As stressed by the above quoted 
authors this photometric offset may be attributed to an additional light 
pattern caused by internal reflections of the telescope corrector. Such 
a pattern, with an amplitude that depends also on the exposure time, is 
present in each image (including flat-fields) and must be corrected 
before flat-fielding. In order to correct for this effect in a proper 
way, one should know the additional-light pattern.

The photometric offset appears to be slightly stronger in the medium-band 
filters than in the broad-band ones (see Figure~\ref{scatterlight} 
right panel). We find an offset in the flux of the stars which may introduce 
an average uncertainty of $\pm$3\% in the broad-band $B$, $V$, $R$, $I$ 
filters and of $\pm$5\% in the medium-band filters.

\subsection{Spectroscopy}

The MOS data reduction was performed using the MIDAS package following 
standard reduction techniques for multi-object slit spectroscopy. For 
each configuration or mask, the reduction consisted of the bias
subtraction, definition of slit positions, flat-fielding, row-by-row 
wavelength calibration, background subtraction, extraction of 
one-dimensional spectra and relative flux calibration. 

For each mask or field, the reduction steps were the following:

\noindent
{\it i)} the average bias frame was subtracted from all the frames 
and a master flat-field was determined using the median option within 
MIDAS;

\noindent
{\it ii)} the definition of the slit positions (along the direction 
perpendicular to the dispersion) and the flat-fielding were performed 
using the master flat and the context MOS within the MIDAS environment;

\noindent
{\it iii)} the three multi-spectra frames of each field were combined 
using a sigma clipping algorithm for the rejection of cosmic ray hits; 

\noindent
{\it iv)} for each configuration or mask, the individual two-dimensional 
long-slit spectra were extracted from the combined multi-spectra 
frames. The next steps were then performed following the standard 
pipeline of the context "long" within the MIDAS package;

\noindent
{\it v)} a row-by-row wavelength calibration was done on each individual 
two-dimensional long-slit spectrum using its corresponding HeAr 
comparison spectrum, extracted exactly in the same way as the 
science spectrum;

\noindent
{\it vi)} the background subtraction was then performed and the 
one-dimensional spectra were extracted;

\noindent
{\it vii)} the spectra of the flux standards, reduced in the same way as 
the science spectra, were then used for the determination of a mean response 
function. A response function was determined for each one of the two
observing nights. Finally, a relative flux calibration was then performed
on the one-dimensional spectra by applying the corresponding response 
function. 

\section{Source extraction from the WFI images}

Catalog extraction was performed using SExtractor (Bertin and Arnouts 1996).  
In order to "increase'' the signal--to--noise ratio, the images are filtered 
with a kernel that, in our case, is a Gaussian with a constant FWHM that 
properly matches the seeing on the images. The source detection is performed 
on the filtered image and the detection threshold is determined as the best 
compromise between limiting magnitude and percentage of spurious detections.  

The photometry was also performed with SExtractor. Magnitudes 
were measured both in fixed circular or {\small CORE} apertures 
(2$\sigma$ and 3$\sigma$ of the PSF FWHM) and in Kron-like (1980) 
elliptical apertures ({\small MAG\_AUTO} in SExtractor, cf. Bertin 
and Arnouts 1996).

{\small MAG\_AUTO} provides the best measurement of the object total flux
because the elliptical aperture is adaptively determined from the object
dimensions. However, in the case of blended objects {\small MAG\_AUTO} does 
not provide reliable results. In these cases (indeed representing a very low 
percentage of the total catalogue), we use the {\small CORE} aperture 
photometry.

We also use the {\small MAG\_AUTO} magnitudes to measure objects colors,
except in the case of blended sources in which {\small CORE} aperture
magnitudes are used.

\begin{table*}
\begin{flushleft}
\caption[]{\label{tab:limits} Completeness magnitudes vs. wavelengths.}
\begin{tabular}{lccccccccc}
\hline
S/N       &$B_{AB}$&$V_{AB}$&$R_{AB}$ &   753 &	770 &  791 & 816 &  837 &  915\\
          &      &     &      &	      &	    &      &     &      &	\\
\hline
10  &  24.6  &   24.0  &   24.3 &  22.8 &   22.4 &   22.1 &   22.5 &   21.8 &  21.9	\\
~5  &  25.3  &   24.8  &   25.1 &  23.7 &   23.3 &   23.0 &   23.4 &   22.7 &  22.8	\\

\hline
\end{tabular}
\end{flushleft}
\end{table*}

\subsection{Completeness magnitude limit of the OACDF}

The best parameter set for the catalog extraction were determined, 
for each final stacked image, using a series of Monte Carlo simulations. 
The goal was: {\it i)} to determine the completeness limit of the 
extracted catalogs, and {\it ii)} to estimate of the fraction of 
spurious detections (see Arnaboldi et al. 2002 for details).

\begin{figure}  
\vspace{8.0cm}
\includegraphics{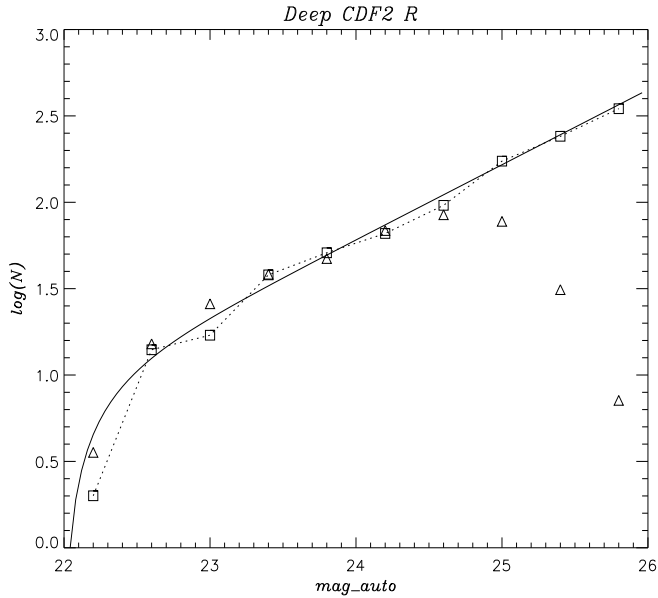}
\caption[]{\label{maglim} Magnitude distributions in the $R$-band mosaic. 
  The simulated sample, connected by the dotted line, is represented with 
  squares, while the extracted sample is represented with triangles. 
  The solid line represents the magnitude distribution of the modeled 
  sample.}
\end{figure}

We ran a series of simulations by randomly adding a synthetic
population of point-like sources with a known luminosity fuction (LF) 
and positions. The completeness limit, i.e. the magnitude at which 50\% 
of the input catalog is lost, was then recovered by matching the
extracted catalog with the modelled population. This procedure
allowed us to derive the completeness limit of $25.1$ AB mag for
the $R$-band mosaic (see Figure~\ref{maglim}). The completeness
limits for the OACDF at S/N ratios of 10 and 5 in the different
bands are listed in Table~\ref{tab:limits}.

In order to estimate the fraction of spurious detections we
proceeded as follows: a simulated image was built from the
background interpolated by SExtractor, by adding the noise map
constructed by multiplying the SExtractor RMS map by a Gaussian
noise image with null mean and $RMS=1$. For a given SExtractor
parameter set, the number of spurious detections was then
determined as the difference between the SExtractor catalog and
the matched catalog with the input modelled population. For the
OACDF $R$-band mosaic, we find that the number of spurious
sources, relative to the total number of detected objects, is less
than 2\% at the completeness limit of 25.1 AB mag.

Besides the usual problems posed by the
bright (saturated) stars which produce a large number of spurious
objects in their extended halos, mosaic images pose additional
problems in the regions at the edges of the individual images
which have lower S/N due to the dithering. The usual approach
would have been to mask out these regions, keeping track of their
size and position to correct the final statistics.
However, so as not to lose completely the information on the
objects falling in such regions, a flag  was introduced in the 
catalogues as follows: we flag as ``BAD'' those objects having 
either high background values (i.e. objects in proximity of saturated 
objects) or high background RMS values (i.e. mostly border objects). 
The threshold to activate such flag was optimized on a trial and 
error basis via the accurate inspection of objects flagged as spurious. 
Extensive testing showed that the best compromise was achieved by 
setting such treshold at either 5 times the value of the local 
background images and/or at a local RMS 20\% larger than the average
RMS of the mosaic itself. These criteria assume that the catalogs 
have been extracted from background subtracted images.

\subsection{Star--galaxy separation}

To distinguish point-like sources from extended ones we use the 
SExtractor parameter {\small CLASS\_STAR}. 
Star/galaxy classification in SExtractor is made using neural networks 
trained on simulated galaxies on single CCD images and is mainly based on 
the deviations of extended objects from a well behaved PSF. 
We use different {\small CLASS\_STAR} indexes for the different bands. 
We set {\small CLASS\_STAR} equal to 0.95, 0.98 and 0.98 for the $B$, $V$ 
and $R_C$ bands respectively.

In mosaic images, the determination of the final PSF is usually non trivial 
due to the combined 
effects of ditherings and varying PSF among the single images. In order to 
test the reliability of the SExtractor {\small CLASS\_STAR} parameter we use 
the spectra obtained for the sample of 173 objects (114 galaxies and 59 stars). 
In this sampe, the objects classified as stars by SExtractor are 63 with 
5\% contamination due to misclassified galaxies, while the objects classified 
as galaxies are 110 with no contaminants. Overall, the SExtractor 
classification turns out to be quite reliable in the magnitude range of 
the spectroscopic sample, ie. brighter than I=22. 

\begin{table}
\begin{flushleft}
\caption[]{\label{tab:counts} Source counts of the OACDF at each band.}
\begin{tabular}{cccc} \hline 
 AB mag.  &   logN$_B$  deg$^{-2}$   &  logN$_V$   deg$^{-2}$  & logN$_R$ deg$^{-2}$     \\
          &                          &                         &                         \\ \hline
  17.0    &  0.90$^{+0.23}_{-0.53}$	 &  1.10$^{+0.19}_{-0.33}$ & 1.46$^{+0.13}_{-0.19}$  \\
  17.5    &  1.00$^{+0.20}_{-0.37}$	 &  1.38$^{+0.13}_{-0.19}$ & 1.72$^{+0.10}_{-0.13}$	 \\
  18.0    &  1.35$^{+0.14}_{-0.22}$	 &  1.74$^{+0.10}_{-0.13}$ & 2.11$^{+0.07}_{-0.08}$	 \\
  18.5    &  1.60$^{+0.11}_{-0.16}$	 &  2.10$^{+0.07}_{-0.08}$ & 2.23$^{+0.06}_{-0.07}$	 \\
  19.0    &  1.77$^{+0.09}_{-0.12}$	 &  2.26$^{+0.06}_{-0.07}$ & 2.56$^{+0.04}_{-0.05}$	 \\
  19.5    &  2.05$^{+0.07}_{-0.08}$	 &  2.62$^{+0.04}_{-0.04}$ & 2.78$^{+0.03}_{-0.03}$	 \\
  20.0    &  2.26$^{+0.06}_{-0.07}$	 &  2.81$^{+0.03}_{-0.03}$ & 2.99$^{+0.03}_{-0.03}$	 \\
  20.5    &  2.42$^{+0.06}_{-0.06}$	 &  3.00$^{+0.02}_{-0.03}$ & 3.17$^{+0.02}_{-0.02}$	 \\
  21.0    &  2.59$^{+0.04}_{-0.04}$	 &  3.17$^{+0.02}_{-0.02}$ & 3.32$^{+0.02}_{-0.02}$	 \\
  21.5    &  2.86$^{+0.03}_{-0.03}$	 &  3.30$^{+0.02}_{-0.02}$ & 3.48$^{+0.02}_{-0.01}$	 \\
  22.0    &  3.05$^{+0.02}_{-0.02}$	 &  3.49$^{+0.01}_{-0.02}$ & 3.62$^{+0.01}_{-0.01}$	 \\
  22.5    &  3.29$^{+0.02}_{-0.02}$  &  3.65$^{+0.01}_{-0.01}$ & 3.77$^{+0.01}_{-0.01}$	 \\
  23.0    &  3.54$^{+0.01}_{-0.01}$	 &  3.86$^{+0.01}_{-0.01}$ & 3.93$^{+0.01}_{-0.01}$	 \\
  23.5    &  3.79$^{+0.01}_{-0.01}$	 &  4.08$^{+0.01}_{-0.01}$ & 4.10$^{+0.01}_{-0.01}$	 \\
  24.0    &  4.03$^{+0.01}_{-0.01}$	 &  4.26$^{+0.01}_{-0.01}$ & 4.24$^{+0.01}_{-0.01}$	 \\
  24.5    &  4.21$^{+0.01}_{-0.01}$	 &  4.37$^{+0.01}_{-0.01}$ & 4.34$^{+0.01}_{-0.01}$	 \\
  25.0    &  4.31$^{+0.01}_{-0.01}$	 &  4.35$^{+0.01}_{-0.01}$ & 4.37$^{+0.01}_{-0.01}$	 \\
  25.5    &  4.41$^{+0.01}_{-0.01}$	 &  4.06$^{+0.01}_{-0.01}$ & 4.34$^{+0.01}_{-0.01}$	 \\ \hline
\end{tabular}
\end{flushleft}
\end{table}

\begin{figure} 
\vspace{19.0cm}
\includegraphics{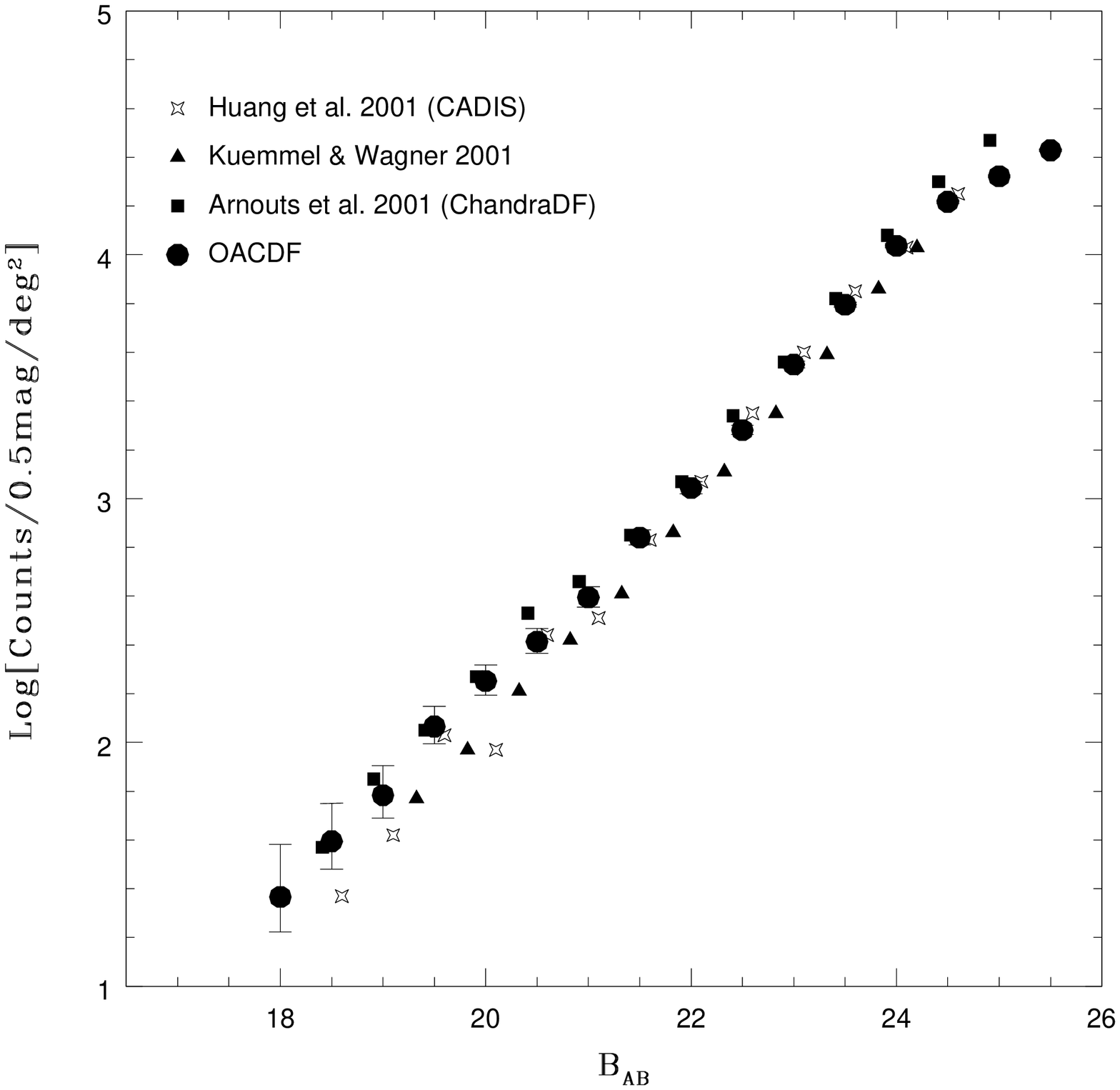}
\includegraphics{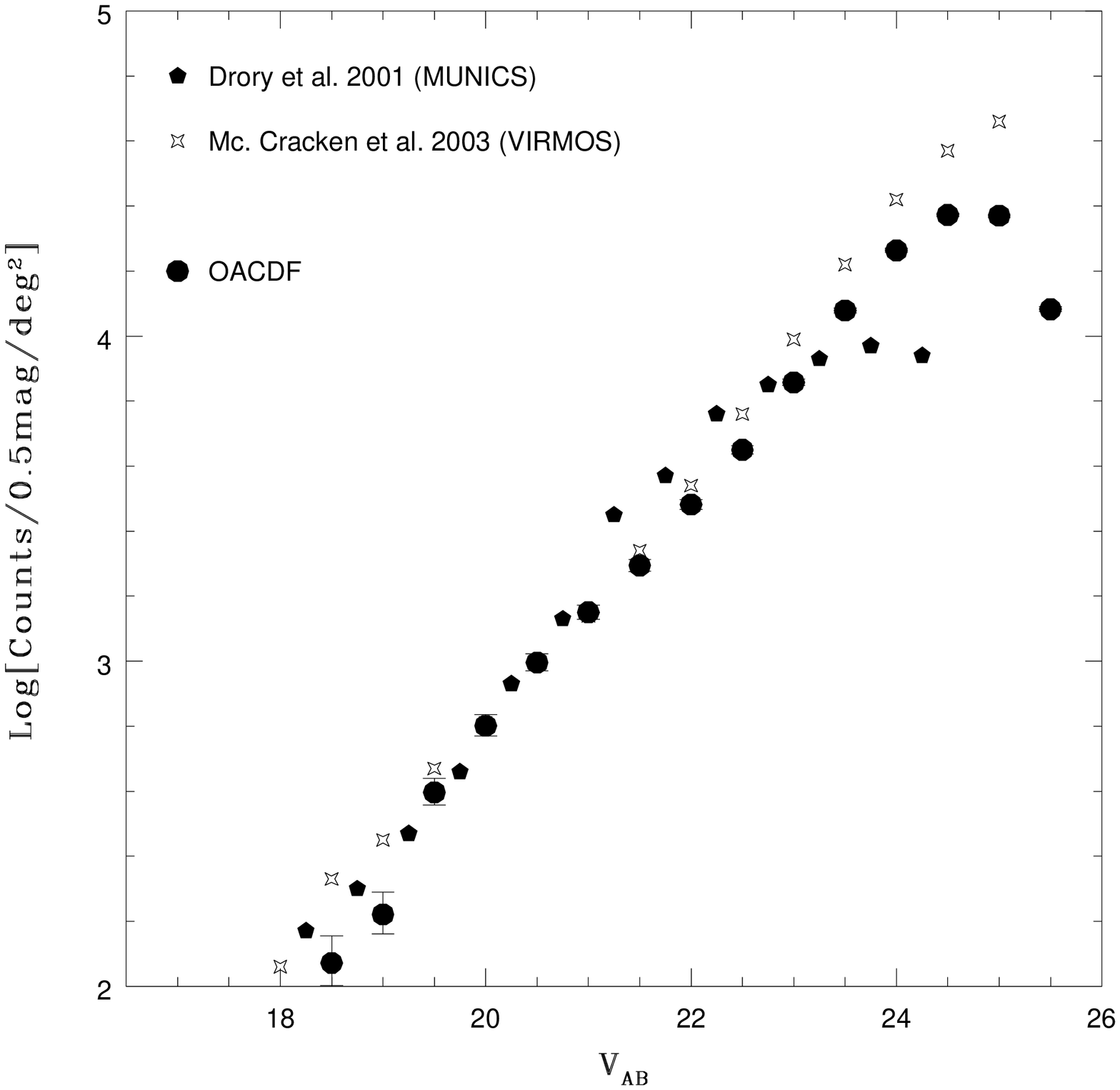}
\includegraphics{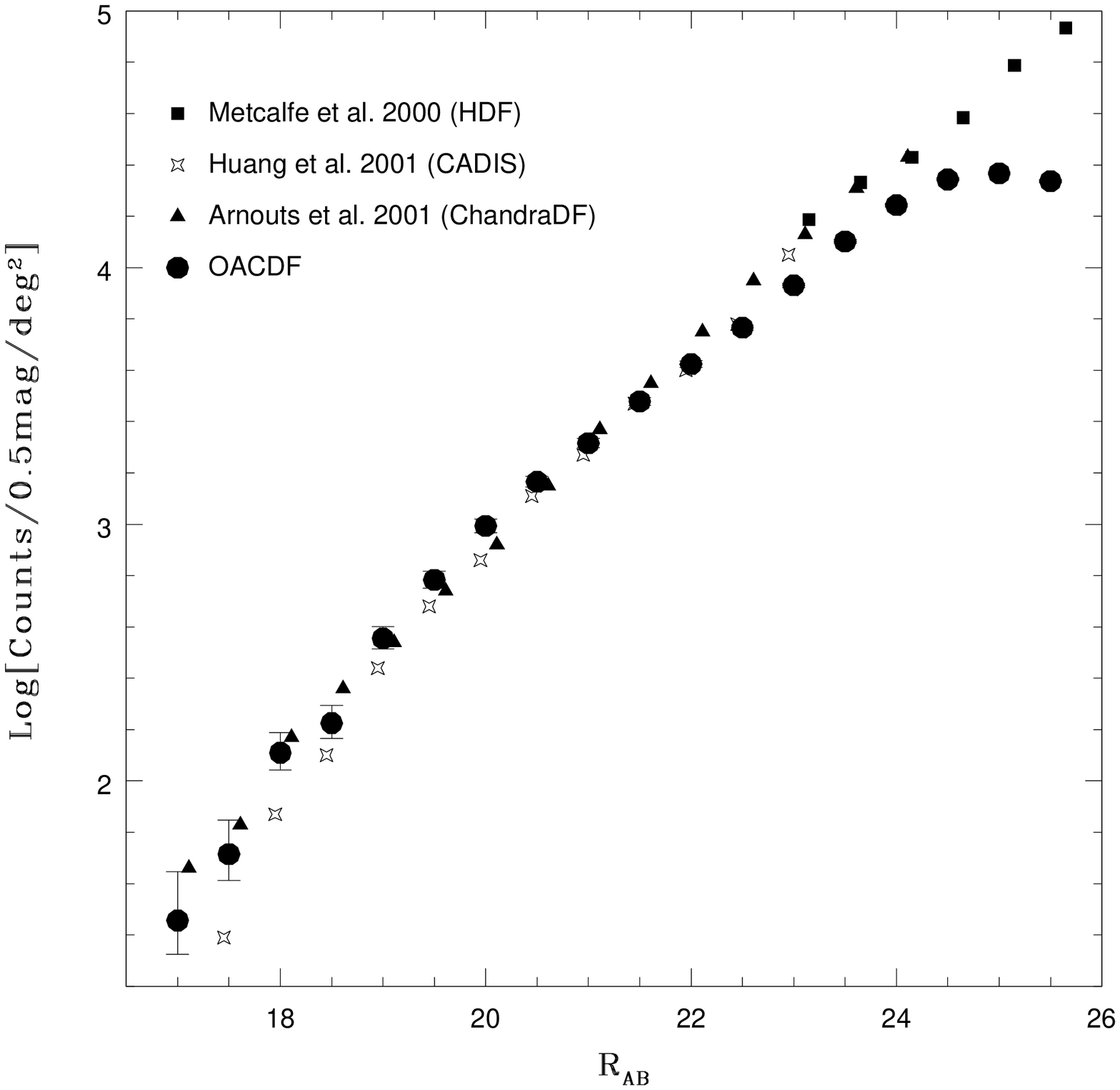}
\caption[]{\label{counts} Comparison of the source number counts in the 
  OACDF (filled circles) with those from the literature. The logarithmic 
  counts are in half-magnitude bins, in the $B$, $V$, and $R$ bands (top, 
  middle and bottom panels respectively) and are normalized to the OACDF deep 
  survey area of 0.5 deg$^2$. The 1$\sigma$ Poissonian errors, also provided 
  in Table~\ref{tab:counts}, are shown.}
\end{figure}

\section{Quality Assessment}

In the following sub-sections we assess the quality of the OACDF catalogs 
by the comparison of the galaxy number counts with those in the literature 
and using the spectroscopic observations of stars and galaxies in the 
OACDF. 

\subsection{Source counts}

In Figure~\ref{counts} the number counts of galaxies in the OACDF, in 
the $B$, $V$, $R_C$ deep fields, are shown. Other number counts from the 
recent literature are also overplotted for comparison. The star/galaxy 
separation is done here on each single catalog by using the stellarity 
index from SExtractor, but without completeness correction applied. 
Despite this, there is a good agreement between the source number 
counts of the OACDF data and those from the literature.

The number counts (log N/0.5mag/deg$^{-2}$) per magnitude bin at each 
band are reported in Table~\ref{tab:counts}. The magnitudes are AB 
magnitudes. The logarithmic counts are reported in half-magnitude bins, 
selected in the $B$, $V$, and $R$ bands, normalized to the effective area of 
the OACDF deep survey (0.5 deg$^2$). 

\subsection{Spectroscopic redshifts}

In order to determine spectroscopic redshifts the Ca II at $\lambda$3933.68 
\& $\lambda$3968.49~\AA\ lines and the Mgb ($\lambda$5173~\AA) absorption 
features, as well as the ``G'' band at $\lambda$4300~\AA\ 
were identified and fitted with Gaussians for the central wavelength 
measurements. In order to further confirm the redshift determinations, a
cross-correlation analysis was applied to the sample of galaxy spectra. 
For this purpose we used the {\em fxcor} task under {\em IRAF}. 
Several late type stars, obtained with the same instrumental set up,
were used as templates for the cross-correlation analysis. Moreover, 
only parts of the spectra, not affected by telluric absorption lines, 
were considered. The typical error of the spectroscopic redshifts is
around $\pm$0.005.
 
The total number of useful spectra is 173. The coordinates, AB magnitudes,
redshifts and spectral types (see next Section) of these objects are 
listed in Tables~\ref{tab:brmags}, ~\ref{tab:nrmags}  and ~\ref{tab:rsh}.
59 of these spectra resulted to be stellar objects.  
The remaining 114 objects are galaxies. For 24 of these, the  redshift 
determination is uncertain because their S/N ratio is less than 10. 
We distinguish these objects in Table~\ref{tab:rsh} with a question mark 
next to the redshift value.

\subsection{Stellar spectral types}

The stellar spectral types were determined by visual comparison of the stellar
spectra with those of the grid of spectral standard stars taken from the 
library of stellar spectra by Jacoby et al. (1984). The most important 
spectral features considered for the spectral type classification were: 
Mg~I $\lambda$5167, $\lambda$5173, and, whenever possible, the ``G'' band 
$\lambda$4300 for the G stars, the Ca~I $\lambda$6162 and Fe~I $\lambda$5227 
for the late G and early K stars, the Na~I $\lambda$5893 and the Fe~I blend 
$\lambda$6495 for the K stars, and the TiO bands $\lambda$4954, $\lambda$5448, 
$\lambda$6159 for the late K and the M stars. 

The magnitudes and spectral types of these stars are provided in 
Tables~\ref{tab:brmags}, \ref{tab:nrmags} and \ref{tab:rsh}. 
In order to transform the $B_{AB}$, $V_{AB}$, $R_{AB}$ and $I_{AB}$ 
magnitudes, reported in Table~\ref{tab:brmags}, into the 
Johnson-Cousins system, the transformation equations 9, 10, 11 
and 12 given in the appendix, must be applied.

In Figure~\ref{stars} the $(B-V)$ and $(V-I_{C})$ colors of the stars 
are plotted versus their spectral types. In these plots the relation between 
color versus spectral type, taken from the stellar library by 
Lejeune et al. (1997), is over-plotted as a continuous line. 
As can be seen, there is a good agreement of the spectral types and colors with 
those of the library of stellar spectra by Lejeune et al. (1997).

\begin{figure} 
\vspace{6.5cm}
\includegraphics{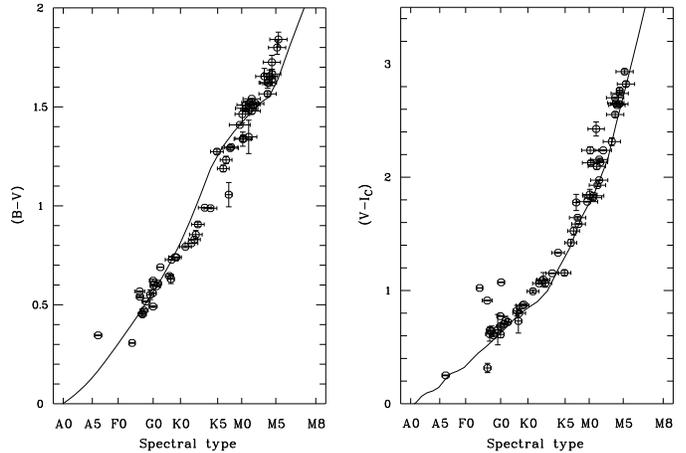}
\caption[]{\label{stars} $(B-V)$ and $(V-I_{C})$ colors of the stars 
  observed with EMMI-MOS versus their spectral type (circles in the left and 
  right panels respectively). The continuous line represents the relationship 
  between the corresponding colors and spectral type, taken from the stellar  
  library by Lejeune et al. (1997).}
\end{figure}

\begin{figure} 
\vspace{8cm}
\includegraphics{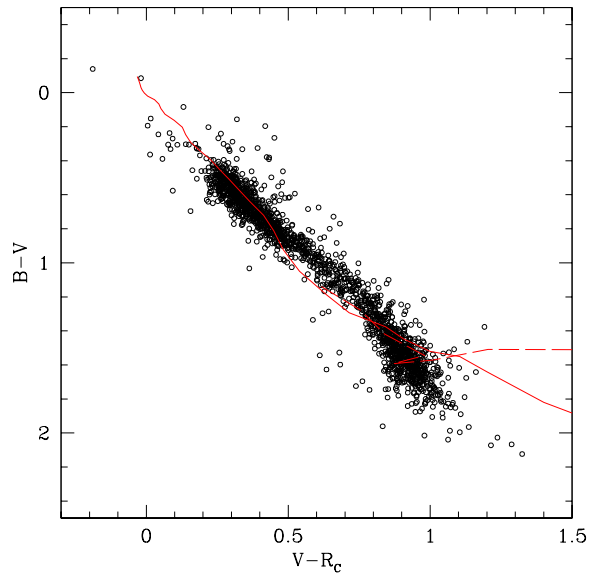}
\caption[]{\label{colcol} Color-color diagram for the objects classified 
 as stars (open circles) using the criterium {\small CLASS\_STAR} $>$ 0.85 
 for the sources detected in the $B$,$V$ and $R$ bands. The synthetic colors 
 for  dwarfs and giants, taken from the standard stellar library by 
 Lejeune et al. (1997), are plotted with the continuous and dashed lines 
 respectively.}
\end{figure}

As a further check, in Figure~\ref{colcol} we show the $(B-V)$ vs. $(V-R_{C})$ 
diagram of point-like sources ({\small CLASS\_STAR} $>$ 0.85) in the OACDF. 
As expected, the colors are consistent with those expected for stars, which
provides a further photometric check of the OACDF catalogs.

\subsection{Photometric redshifts}

Photometric redshifts were determined using the HYPERZ package 
(see Bolzonella, Miralles and Pell\'o 2000 for details). The spectroscopic 
redshifts were used to fine-tune the software parametrization. The optimal 
set-up, which yields the most consistent results with the spectroscopic 
redshifts, was to use three broad-bands ($B$, $V$, $R$) plus the medium-band 
filters centered at 753 and 914~nm. We use the {\small MAG\_AUTO} magnitudes 
(see Section~4) for the photometric redshifts computation because they 
provide more robust results for all type of objects in the catalog. 

The procedure we used for the photometric redshift determinations does 
not include templates for QSOs. Therefore, the results for objects like 
OACDF122436.7-124519, which is a relatively high-redshift QSO ($z \sim$3), 
are also unreliable. For this particular case, 
HYPERZ provides a photometric redshift of z$_{phot}$=0.23. In 
forthcoming papers, we will include AGN--like templates in order to 
minimize such large residuals.

\begin{figure} 
\vspace{6.5cm}
\includegraphics{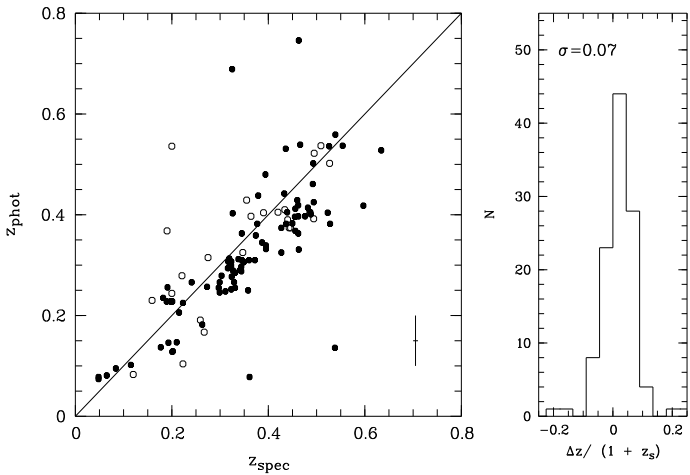}
\caption[]{\label{fotreds}
 Left panel: photometric versus spectroscopic redshifts for the objects 
 reported in Table~\ref{tab:rsh}. The black dots represent the galaxies 
 with well determined spectroscopic redshifts, while the open circles 
 represent the 24 galaxies with dubious redshift determinations (indicated 
 with a ``?'' next to $z_{spec}$ in Table~\ref{tab:rsh}). 
 The continuous line represents the relation z$_{phot}$ = z$_{spec}$; right 
 panel: distribution of the difference z$_{spec}$ - z$_{phot}$.
}
\end{figure}

The spectroscopic redshifts versus the photometric redshifts of different 
types of objects in the OACDF2 are plotted in Figure~\ref{fotreds}. 
The black dots represent the galaxies with well determined spectroscopic 
redshifts, while the open circles represent the 24 galaxies with dubious 
redshift determinations (indicated with a question mark next to the 
spectroscopic redshift in Table~\ref{tab:rsh}).

\begin{figure} 
\vspace{8cm}
\includegraphics{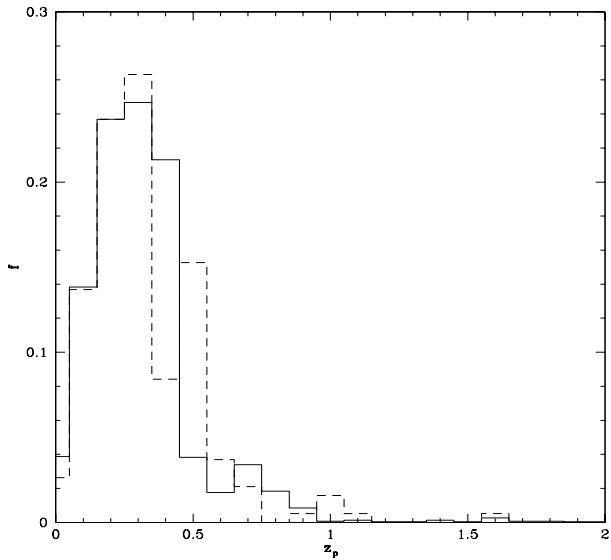}
\caption[]{\label{fotreds_distr} Distribution of photometric redshifts 
 of the high S/N OACDF catalog (continuous line) compared with the redshift 
 distribution of the CFRS (dashed line) at the same depth.}
\end{figure}

Despite the dubious redshifts, the vast majority of the photometric redshifts 
match well the spectroscopic redshifts within the errors. Six out of 113 
objects fall off the general trend (a percentage compatible with that 
of previous works). We find that the mean difference between photometric and 
spectroscopic redshifts is 0.02$\pm$0.07 (see Figure~\ref{fotreds} right 
panel). 
Therefore, our photometric redshift determinations are confident within an 
RMS error of about 0.07, in the redshift range from 0 to 0.7. 

The distribution of photometric redshifts of the OACDF, is shown in 
Figure~\ref{fotreds_distr}. The distribution of the CFRS (Canada France 
Redshift Survey, Lilly et al. (1995)) is over-plotted with a dashed line 
for comparison.

For the sake of this comparison, we have selected two classes of objects 
from the whole catalog: 
{\it i)} the bright class is composed by objects with a high photometric 
S/N ratio and hence, objects for which the results obtained from the 
comparison with the spectroscopic follow-up are consistent; 
{\it ii)} the faint class, i.e. all the objects fainter than $V_{AB}=$~22~mag. 
Although the latter were detected with a S/N $>$ 10, we still flagged them 
because we do not have any spectroscopic confirmation of their redshifts. 
The figure shows the objects in the bright class only. 

In order to confirm the similarity of the two redshift distributions
shown in Figure~\ref{fotreds_distr} we performed a Kolmogorov-Smirnov 
two sample test (KS test), which is a non-parametric test that yields correct
probabilities without conditions for the distribution of the errors. 
The result of the KS test is that the two redshift distributions are similar 
at the 99.4\% confidence level.
 We can thus conclude that the two distributions come from the same parent 
distribution.

\section{Search for groups and clusters of galaxies}

An accurate knowledge of the global properties of galaxy clusters and groups 
is needed both to constrain the  models of galaxy formation and evolution and
to falsify the cosmological models. The first step in the study  of galaxy
clusters is, however, the construction of statistically significant and unbiased
catalogues of cosmic structures spanning a wide range of richness (N):
from very low multiplicity structures  such as galaxy triplets (N=3), to
groups (N$<$50), and clusters (N$>$50). While much work has been devoted to 
the construction of complete rich cluster samples, the low multiplicity end
of the structure spectrum is still poorly  known due to the difficulties
encountered in identifying physically bound systems of low multiplicity on 
wide field or survey material. On projected data, in fact, loose and poor
groups of galaxies produce low signal/noise overdensities which are
difficult to detect on shallow photometric material and require spectroscopic
surveys and/or deep fields covering moderately wide areas. 

Not so much work has been done to detect poor structures such as loose groups; 
two principal methods (and their successive elaborations) have been adopted. 
In particular, Turner \& Gott (1976) presented the first tentative objective 
identification 
of groups as enhancements above a reliable threshold in the projected galaxy 
distribution. A noticeable exception to the lack of low-richness catalogs 
has been the detection of compact groups, where several teams 
(de Carvalho \& Djorgovski 1995; Iovino et al. 1999, 2003) have proposed 
different approaches to their detection. For the determination of the 
mass function, it is important to note that its derivation from the 
above-cited catalogues is hindered by the fact that all of the above 
algorithms are  optimized for the detection of either groups or clusters, 
and with a few exceptions (cf. Puddu et al. 2003) no systematic work has 
been done in matching their outcomes in the transition region between 
structures of low and high richness. 
The OACDF, as any other deep field, offers the possibility to begin covering 
this gap. In fact, by assuming an Euclidean metric and uniform distribution 
we expect to find in the OACDF ca. 5 groups in the richness range [5,30] 
(against less than one rich cluster). Since the investigated area of the 
OACDF is limited to 0.5 $\deg ^2$, we expect to find about 4 groups in the 
richness range [5,30] and less than one cluster in the richness range 
[30,100], in the volume enclosed within 0.5 $\deg ^2$ and z$<$1.0.

\begin{figure}
\vspace{15.5cm}
\includegraphics{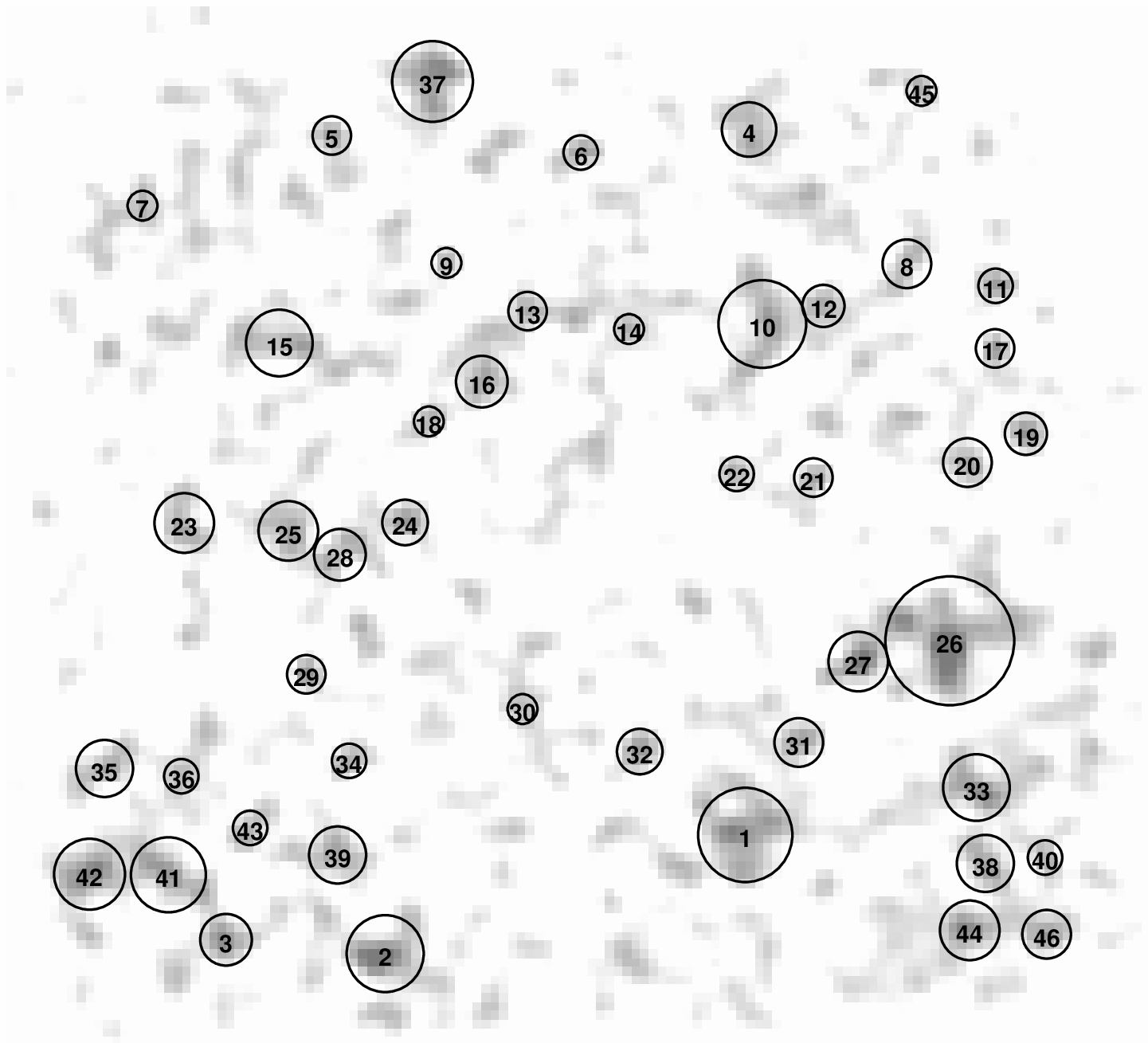}
\includegraphics{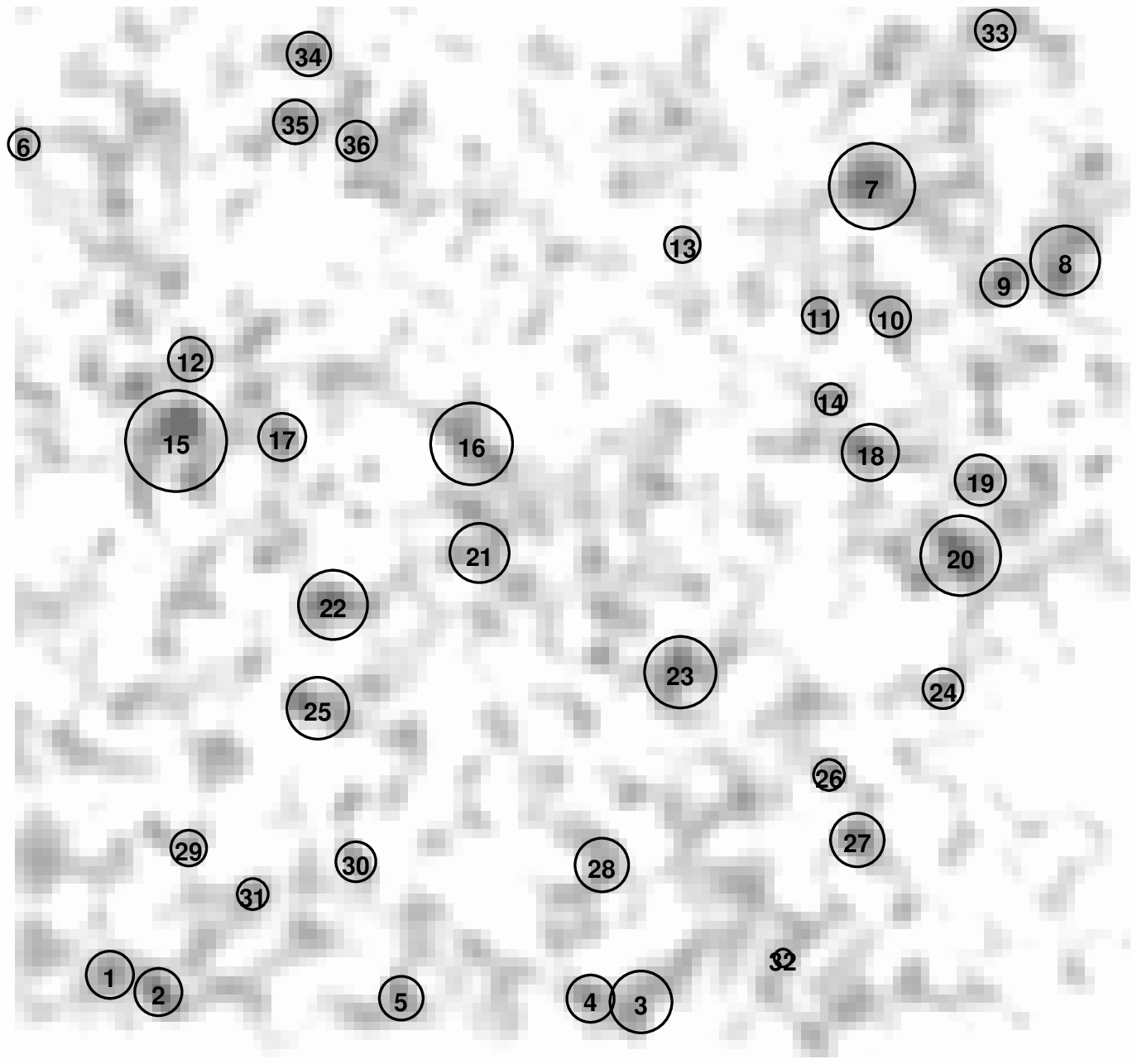}
\caption[]{\label{smomap} Smoothed density maps of the OACDF2 (top) and 
 OACDF4 (bottom) fields. Each map has been produced by binning the matched 
 catalogs (in the $B$, $V$ and $R$ bands) into equal area square bins of 
 $18 \times 18$ squared arc-sec (see Puddu et al. 2003 for details). The 
 circles mark the detected over-densities (including those having a low S/N 
 ratio).} 
\end{figure}

In order to actually detect them we used
the method developed by Puddu et al. (2003) and already applied to the DPOSS 
data (about 300 $\deg ^2$; see Puddu et al. 2001). 
In summary: we first extracted a catalogue containing all extended
sources  with 20.5 $< R <$ 24 and then binned the corresponding galaxy spatial
distribution into equal area square bins to produce a density map, such
that the mean number of galaxies per bin is $\sim 1$ (binsize=0.005 deg). In
order to identify and extract the density peaks on the resulting density
map, we run SExtractor (after background  independent estimation and
subtraction) with an absolute threshold of 0.5 (objects/pixel) and a minimum
detection area of 4 pixels. 
The background subtraction removes all the structures with spatial frequencies
smaller than the scale length of 50 pixels, which corresponds to linear 
dimensions of $3-4$ Mpc, in the redshift range $0.3, 0.6$. For the background
galaxies, we find a mean  density of $\sim 43000 \pm 4200$ objects per square
degree.  In order to enhance the structures, the detection was ran on the
density map convolved with a Gaussian filter $3$ pixels wide (corresponding
to 150-250 kpc, at redshift $0.3-1.0$). The latter are reliable dimensions 
for a cluster  core. A caveat of our method is that the Gaussian filter sets
strong limitations on the possibility of detecting  structures at higher
redshifts, which appear with a more compact core than the minimum dimension 
of the filter itself.  The over-density regions are defined by the
isodensity contours corresponding to the detection threshold on a  smoothed
map (Figure 8). The area defined by these contours is called the iso-area. We
used the number of  objects lying inside this area to give an estimate of
the cluster candidate richness. The error on the richness estimate depends
on the S/N ratio of the detection and is inversely correlated with the richness
itself: low and intermediate richness ($<$ 30) candidates have a low detection
S/N and hence are not detected with an high  confidence. We then use 
colour-magnitude diagrams to look for a possible red sequence of early type 
galaxies, confirming that the overdensity corresponds to a galaxy cluster.  

The cluster red sequence can be isolated from the background objects 
using a pair of filters, if these filters sample the 4000\AA\ break. 
In our case we chose the $(V-R_C)$ color and, in order to explore a 
larger redshift range, we also use the $V-913$ color. There is, however, 
the caveat that the limiting magnitude in the narrow band filters is 
brighter than for the $V$ and $R_C$ bands with the consequence that not 
all the objects present in the $(V-R_C)$ sample will be retrieved in 
the $V-913$ sample. However, this is not a problem for our analysis 
because we are interested in the bright ellipticals of the red 
sequence rather than in faint background galaxies. 

\begin{figure} 
\vspace{8cm}
\includegraphics{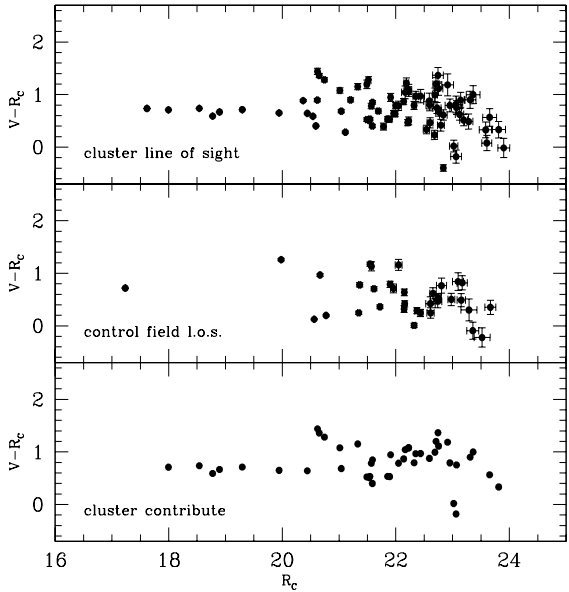}
\caption[]{\label{ssub} Example of the statistical subtraction applied 
 to the cluster candidate OACDF2\_2. The $(V-R_C)$ vs. $R_C$ diagram for 
 all the objects (cluster plus background) in the over-density region is 
 shown in the upper panel, while the same diagram for the background objects 
 is shown in the middle panel; the results after the statistical 
 subtraction are shown in the bottom panel.}
\end{figure}

For each over-density, we plot the color-magnitude diagrams $(V-R_C)$  vs. 
$R_C$ and $V-913$ vs. 913 of all the objects in an aperture with an area 
equivalent to three times the iso-area and centered on the over-density 
centroid. We then select a comparison field in an annular region centered 
on the cluster, with a much larger radius, but with comparable area as
that defined by the over-density. 
We then perform a statistical subtraction of the background contribution, 
which consists on the elimination of the points nearest to the neighbor 
background points in the color-magnitude diagram. In this way, the red 
sequence is enhanced, if it is really present.
As an example, we show in Figure~\ref{ssub} the procedure applied to the 
cluster candidate OACDF-CLG02. 

\begin{table*}
\begin{flushleft}
\caption[]{\label{tab:clusters} Candidates for clusters of galaxies}
\begin{tabular}{lcccccc}
\hline
candidate     & RA(2000)     & DEC (2000)     & richness & iso-area & S/N & ph\_z \\
              &              &                &          & (pixs)   &	    &       \\ \hline
OACDF-CLG01     & 12 \ 23 \ 58 & -12 \ 44 \ 53  & 59	& 30	   & 5.7 & 0.3   \\
OACDF-CLG02$^*$ & 12 \ 24 \ 47 & -12 \ 48 \ 56  & 49	& 20	   & 5.3 & 0.2   \\
OACDF-CLG03    & 12 \ 23 \ 56 & -12 \ 27 \ 29  & 36	& 26	   & 4.3 & 0.5   \\
OACDF-CLG04    & 12 \ 25 \ 00 & -12 \ 34 \ 32  & 22	& 12	   & 3.4 & 0.3   \\
OACDF-CLG05    & 12 \ 25 \ 27 & -12 \ 46 \ 14  & 25	& 17	   & 3.7 & 0.3   \\
OACDF-CLG06    & 12 \ 23 \ 32 & -12 \ 53 \ 20  & 32	& 23	   & 4.1 & 0.3   \\
OACDF-CLG07    & 12 \ 25 \ 04 & -13 \ 01 \ 41  & 65	& 32	   & 6.1 & 0.2   \\    
OACDF-CLG08    & 12 \ 24 \ 25 & -13 \ 01 \ 47  & 28	& 21	   & 3.9 & 0.3   \\
OACDF-CLG09    & 12 \ 23 \ 18 & -13 \ 02 \ 58  & 17	&  8	   & 3.3 & 0.1   \\
OACDF-CLG10    & 12 \ 23 \ 21 & -13 \ 05 \ 27  & 46	& 20	   & 5.2 & 0.2   \\ \hline

*: This cluster has been spectroscopically confirmed. 
\end{tabular}
\end{flushleft}
\end{table*}
 
\begin{figure*} 
\vspace{14cm}
\includegraphics{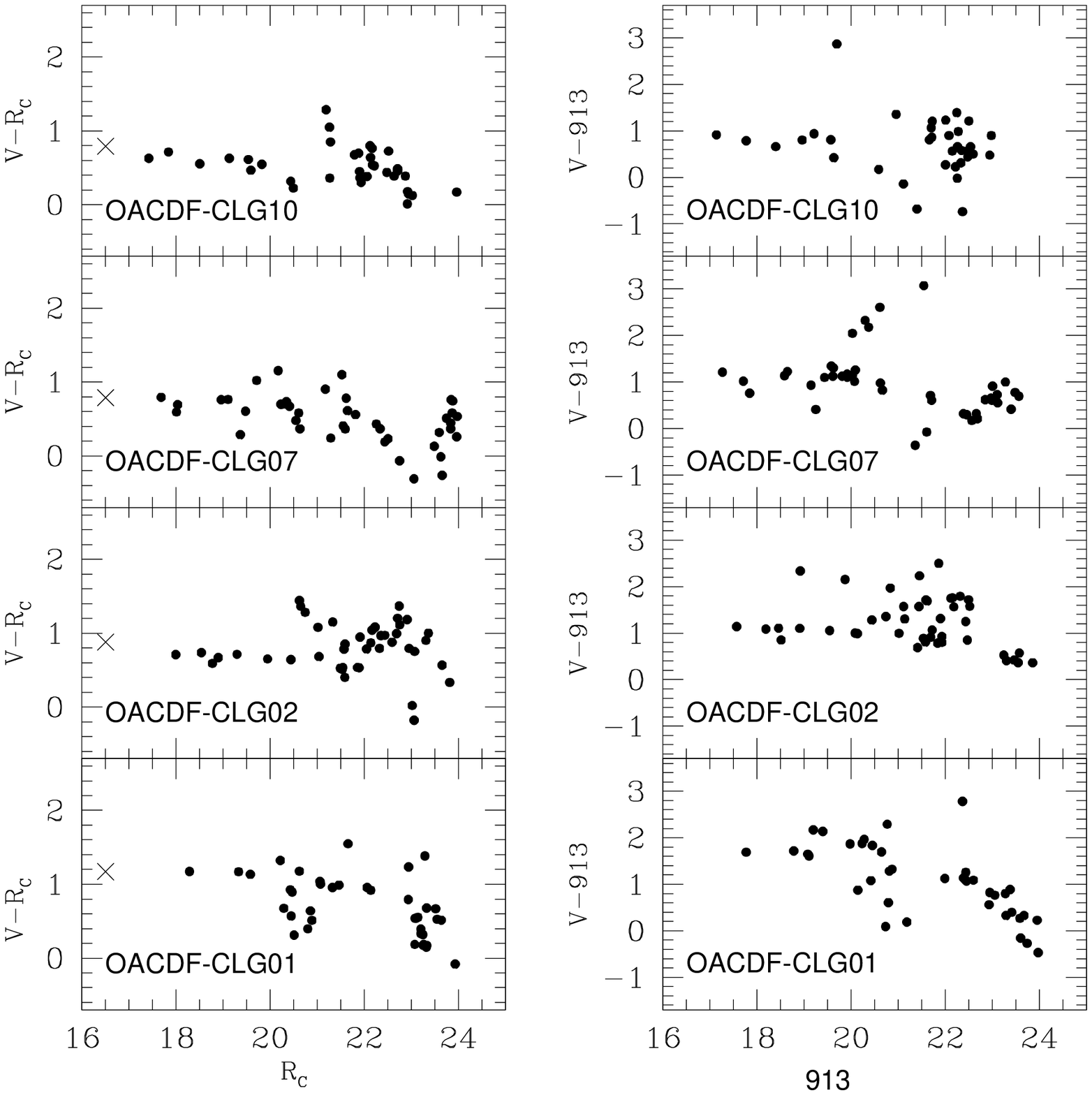}
\caption[]{\label{CMocl} Color -- magnitude diagrams of our four best 
 candidates, after applying  the statistical subtraction of the 
 background objects. The "$\times$" symbol on the (V-R$_C$) vs. R$_C$ diagrams 
 represents the color corresponding to the photometric redshift reported 
 in Table~\ref{tab:clusters}.}
\end{figure*}

Ten candidates for clusters of galaxies were selected.  
However, we note that a large number of low S/N over-densities, 
that coincide with many groups of galaxies, is detected in the
OACDF.
In Figure~\ref{CMocl} we plot the $(V-R_C)$ vs. $R_C$ and $V-913$ vs. 913 
color-magnitude diagrams of our four best cases selected on the OACDF2 
and OACDF4 after the statistical subtraction of the background objects 
on the color-magnitude plane. 
 
The coordinates, estimated richness, iso-area (in pixels), S/N ratio
and the estimated photometric redshift for a list of galaxy cluster 
candidates, detected with S/N$>$3 and selected on the basis of their 
red sequences, is provided in Table~\ref{tab:clusters}. The coordinates 
correspond to the over-density centroids. The richness estimate was done 
as described above. The photometric redshift for each candidate
was estimated as the median of the photometric redshifts of the objects 
in the corresponding aperture (three times the iso-area).
 
It is interesting to note that the color-magnitude diagram of the cluster
candidate OACDF-CLG02 shows some hint of a double red sequence, which 
is more evident in the $V-913$ vs. 913 diagram. This leads to the idea of 
two galaxy clusters at different redshifts (one at z=0.23 and the other 
one at z$\approx$0.6) which by chance projection fall approximately along 
the same line of sight. The spectroscopic observations reported in Section~5
confirm the first galaxy cluster at z=0.20. This cluster is also known as 
LCDCS 0522, and was first discovered in the Las Campanas Distant 
Cluster Survey (Gonzalez et al. 2001), which estimated the redshift to 
be $z \simeq 0.34$. Given the uncertainty of about 0.1 in the latter
redshift, and considering that the redshift of this cluster was confirmed
spectroscopically by us using spectra of seven galaxies in the cluster,
we take as correct our redshift determination. 
%

\section{Summary}

The OACDF, which comprises a photometric and astrometric database 
for galactic and extragalactic research, has been presented. The
observations (imaging and spectroscopy), data reduction and photometric 
calibration, as well as the source extraction techniques, were 
descrived.

During the course of the observations with the WFI at the ESO 2.2m
telescope, a photometric inhomogeneity was found that may be attributed, 
at least in part, to an additional-light pattern due to internal 
reflections off the telescope corrector, that may yield photometric 
errors of up to 3\% in the $B$, $V$, $R$ and $I$  broad-bands and up 
to 5\% in the medium-band filters. 

The quality assessment studies of the OACDF provide the following results:
the galaxy number counts of the OACDF deep data, in the $B$, $V$ and 
$R$ bands, are consistent with those from the literature; the resulting 
photometric redshifts, for objects brighter than about $V=$23, are consistent 
with the spectroscopic redshifts within a RMS error of about 0.07 in the 
redshift range from 0 to 0.7; the distribution of photometric redshifts 
of OACDF sources is consistent with that of the CFRS at the 99.4\% confidence 
level. The relationship between observed stellar colors and stellar spectral 
types for the 59 stars in our spectroscopic sample fits well the intrinsic 
color versus spectral type relation in the spectral range from A5 to M5. 
This confirms the low interstellar extinction of these stars, and 
hence the low galactic extinction in the OACDF.

We therefore conclude that the depth of the OACDF and its relatively large 
spatial coverage with respect to pencil beam surveys make it a good tool 
for extragalactic and stellar studies. In particular it is well suited for 
the definition of a statistically significant sample of early type galaxies 
for subsequent studies of galaxy formation and evolution in the redshift 
range 0--1.

The OACDF also comprises a suitable database for the search for groups
of galaxies and intermediate richness ($N_{gal} <100$) clusters at 
intermediate redshift. Applying our cluster detection algorithms, several 
candidates for clusters of galaxies were detected above the 3$\sigma$ level. 
Using the criterion of the red sequence in the color-magnitude diagram, 
we selected 10 candidates with estimated richness in the range from 
20 to 60 and a photometric redshift in the range from 0.2 to 0.5. One of 
these (OACDF-CLG02) has been spectroscopically confirmed to have a redshift 
of 0.2. The other 9 need spectroscopic confirmation. 
Future studies of these cluster candidates will allow us to improve our 
knowledge 
of the different evolutionary histories of galaxies in different enviroments.
The multiband and sky coverage of the OACDF make it a suitable survey to 
test the predictions of different theoretical scenarios at intermediate 
and high redshift.

\begin{acknowledgements}

 We thank the anonymous referee for comments and suggetions
 which helped to improve a first version of this paper. We also 
 thank S. Andreon, G. Busarello, E. Covino, D. de Martino, 
 M. Massarotti for discussions. 
 The assistantship of G. Thereau during the first observing run 
 contributed to make those nights more profitable. 
 We are grateful to H. McCracken for discussions regarding a first 
 recipe for the astrometry. We thank Prof. M. Dopita  for comments 
 and suggestions on an earlier version of this paper.
 We also thank E. Bertin for many useful discussions and suggestions 
 regarding catalog extraction methods and F. Valdes for several 
 suggestions on the use of the IRAF mscred package.
 We are grateful to A. Di~Dato, M. Colandrea and K. Reardon 
 for their help with the INAF-OAC computers. Finally, we want to 
 thank the ESO 2.2m telescope team for their assistance during the 
 observations. This project has been partially financed by the 
 Italian {\it Ministero dell'Universit\`a e della Ricerca Scientifica 
 e Tecnologica} (MURST).

\end{acknowledgements}

\appendix

\section{The Catalog}

In this appendix we provide a catalog containing magnitudes and 
spectroscopic information for the 173 objects. The catalog is given 
in Tables~\ref{tab:brmags}, \ref{tab:nrmags} and \ref{tab:rsh}.

The broad-band AB magnitudes reported in the catalog can be transformed into
the Johnson-Cousins system by applying the following  expressions:

 \begin{equation}
  V = 0.870 \cdot V_{AB} + 0.130 \cdot R_{AB} - 0.030
 \end{equation}

 \begin{equation}
  R_{C} = 1.030 \cdot R_{AB} - 0.030 \cdot V_{AB} - 0.221
 \end{equation}

 \begin{equation}
  I_{C} = 0.907 \cdot I_{AB} + 0.093 \cdot V_{AB} - 0.462
 \end{equation}

 \begin{equation}
  B = 1.311 \cdot B_{AB} - 0.311 \cdot V + 0.122
 \end{equation}


\begin{table*}
\begin{flushleft}
\input tab5a.tex
\end{flushleft}
\end{table*}

\begin{flushleft}
\begin{table*}
\input tab5b.tex
\end{table*}
\end{flushleft}

\begin{flushleft}
\begin{table*}
\input tab5c.tex
\end{table*}
\end{flushleft}

\begin{table*}
\begin{flushleft}
\input tab6a.tex

\end{flushleft}
\end{table*}

\begin{flushleft}
\begin{table*}
\input tab6b.tex
\end{table*}
\end{flushleft}

\begin{flushleft}
\begin{table*}
\input tab6c.tex
\end{table*}
\end{flushleft}

\begin{flushleft}
\begin{table*}
\input tab7a.tex
\end{table*}
\end{flushleft}

\begin{flushleft}
\begin{table*}
\input tab7b.tex
\end{table*}
\end{flushleft}

\begin{flushleft}
\begin{table*}
\input tab7c.tex
\end{table*}
\end{flushleft}

\end{document}

%% file: tab5a.tex



\caption[]{\label{tab:brmags} Coordinates and broad-band AB magnitudes for 
           173 objects in the OACDF2}

\small

\begin{tabular}{lcccccc}  \hline

  OACDF         &   RA(2000)   &  DEC(2000)   &     $B_{AB}$   &      $V_{AB}$		&  $R_{AB}$	     &      	$I_{AB}$  \\
                &              &              &                &		              &		              &	                \\
                &              &              &                &		              &		              &	                \\   \hline
                &              &              &                &		              &		              &	                \\
122325.1-123757 & 12:23:25.077 & -12:37:57.33 & 22.72$\pm$0.07 & 22.11$\pm$0.05 & 20.97$\pm$0.03 & 19.95$\pm$0.10  \\
122325.4-123915 & 12:23:25.418 & -12:39:15.02 & 19.98$\pm$0.04 & 19.05$\pm$0.03 & 18.24$\pm$0.03 & 17.06$\pm$0.08  \\
122327.4-123714 & 12:23:27.351 & -12:37:13.67 & 22.00$\pm$0.05 & 21.38$\pm$0.03 & 20.69$\pm$0.03 & 19.94$\pm$0.10  \\
122327.5-123718 & 12:23:27.464 & -12:37:18.25 & 21.86$\pm$0.05 & 21.08$\pm$0.03 & 20.07$\pm$0.03 & 19.22$\pm$0.09  \\
\hline

\end{tabular}

%% file: tab5b.tex
\normalsize

\begin{center}
 {{\bf Table 5.} Continued}
\end{center}

\vspace{0.3cm}

\small

\begin{tabular}{lcccccc}  \hline

  OACDF         &   RA(2000)   &  DEC(2000)   &     $B_{AB}$    &      $V_{AB}$		 &  $R_{AB}$	     &      	$I_{AB}$   \\
                &              &              &                 &		               &		              &	                 \\
                &              &              &                 &		               &		              &	                 \\   \hline
                &              &              &                 &		               &		              &	                 \\
122426.5-123002 & 12:24:26.499 & -12:30:02.32 &  21.98$\pm$0.04 & 21.70 $\pm$0.03 & 21.57$\pm$0.03 & 21.43$\pm$0.13  \\
122426.6-121947 & 12:24:26.610 & -12:19:47.24 &  16.58$\pm$0.04 & 16.41 $\pm$0.03 & 16.60$\pm$0.03 & 16.64$\pm$0.08  \\
122427.4-122313 & 12:24:27.382 & -12:23:13.46 &  22.20$\pm$0.05 & 21.17 $\pm$0.03 & 20.29$\pm$0.03 & 18.50$\pm$0.08  \\
122427.4-122924 & 12:24:27.424 & -12:29:24.21 &  16.79$\pm$0.04 & 16.40 $\pm$0.03 & 16.26$\pm$0.03 & 16.06$\pm$0.08  \\
\hline
\end{tabular}

%% file: tab5c.tex
\normalsize

\begin{center}
 {{\bf Table 5.} Continued}
\end{center}

\vspace{0.3cm}

\small

\begin{tabular}{lcccccc}  \hline

  OACDF         &   RA(2000)   &  DEC(2000)   &     $B_{AB}$    &      $V_{AB}$		&  $R_{AB}$	     &      	$I_{AB}$  \\
                &              &              &                 &		              &		              &	                \\
                &              &              &                 &		              &		              &	                \\   \hline
                &              &              &                 &		              &		              &	                \\
122447.6-124852 & 12:24:47.640 & -12:48:52.32 &  23.21$\pm$0.10 & 22.22$\pm$0.06 & 20.91$\pm$0.03 & 20.05$\pm$0.09  \\
122448.2-122137 & 12:24:48.170 & -12:21:37.21 &  23.61$\pm$0.10 & 23.02$\pm$0.06 & 22.20$\pm$0.04 & 21.43$\pm$0.13  \\
122448.5-122847 & 12:24:48.505 & -12:28:47.13 &  22.16$\pm$0.05 & 21.08$\pm$0.03 & 20.13$\pm$0.03 & 19.45$\pm$0.09  \\
122448.8-124703 & 12:24:48.842 & -12:47:02.94 &  21.52$\pm$0.04 & 20.84$\pm$0.03 & 20.52$\pm$0.03 & 19.97$\pm$0.10  \\
\hline

\end{tabular}

%% file: tab6a.tex



\caption[]{\label{tab:nrmags} Intermediate-band AB magnitudes for 173 objects 
           in the OACDF2}

\small

\begin{tabular}{lccccccc}  \hline

  OACDF         &      M753	     &   M770		       &       M791 	   &    M816	       &       M837	    &      M884 	    &     M913		     \\
                &                &		              &	               &	               &	               & 		             & 		             \\
                &                &		              &	               &	               &	               & 		             & 		             \\   \hline
                &                &		              &	               &	               &	               & 		             & 		             \\
122325.1-123757 & 20.55$\pm$0.07 & 20.55$\pm$0.07 & 20.30$\pm$0.07	& 20.17$\pm$0.06	& 20.09$\pm$0.07 & 20.04$\pm$0.07 & 20.19$\pm$0.07 \\
122325.4-123915 & 17.31$\pm$0.06 & 17.37$\pm$0.06 & 17.15$\pm$0.06	& 17.02$\pm$0.06	& 16.99$\pm$0.06 & 16.93$\pm$0.06 & 17.00$\pm$0.06 \\
122327.4-123714 & 20.32$\pm$0.06 & 20.29$\pm$0.06 & 20.10$\pm$0.07	& 20.02$\pm$0.06	& 20.07$\pm$0.07 & 19.94$\pm$0.06 & 20.06$\pm$0.07 \\
122327.5-123718 & 19.63$\pm$0.06 & 19.52$\pm$0.06 & 19.41$\pm$0.06	& 19.27$\pm$0.06	& 19.21$\pm$0.06 & 19.17$\pm$0.06 & 19.24$\pm$0.06 \\
\hline

\end{tabular}

%% file: tab6b.tex
\normalsize

\begin{center}
 {{\bf Table 6.} Continued}
\end{center}

\vspace{0.3cm}

\small

\begin{tabular}{lccccccc}  \hline

  OACDF         &      M753	     &   M770		       &       M791 	   &    M816	       &       M837	    &      M884 	    &     M913		     \\
                &                &		              &	               &	               &	               & 		             & 		             \\
                &                &		              &	               &	               &	               & 		             & 		             \\   \hline
                &                &		              &	               &	               &	               & 		             & 		             \\
122426.5-123002 & 21.56$\pm$0.07 & 21.43$\pm$0.07 & 21.65$\pm$0.11	& 21.47$\pm$0.07	& 21.39$\pm$0.11 & 21.42$\pm$0.08 & 21.64$\pm$0.09 \\
122426.6-121947 & 16.29$\pm$0.06 & 16.41$\pm$0.06 & 16.47$\pm$0.06	& 16.66$\pm$0.06	& 16.30$\pm$0.06 & 16.30$\pm$0.06 & 16.66$\pm$0.06 \\
122427.4-122313 & 19.02$\pm$0.06 & 19.05$\pm$0.06 & 18.78$\pm$0.06	& 18.38$\pm$0.06	& 18.43$\pm$0.06 & 18.26$\pm$0.06 & 18.23$\pm$0.06 \\
122427.4-122924 & 16.08$\pm$0.06 & 16.08$\pm$0.06 & 15.92$\pm$0.06	& 15.91$\pm$0.06	& 15.90$\pm$0.06 & 15.91$\pm$0.06 & 16.06$\pm$0.06 \\
\hline

\end{tabular}

%% file: tab6c.tex
\normalsize

\begin{center}
 {{\bf Table 6.} Continued }
\end{center}

\vspace{0.3cm}

\small

\begin{tabular}{lccccccc}  \hline

  OACDF         &      M753	     &   M770		       &       M791 	   &    M816	       &       M837	    &      M884 	    &     M913		     \\
                &                &		              &	               &	               &	               & 		             & 		             \\
                &                &		              &	               &	               &	               & 		             & 		             \\   \hline
                &                &		              &	               &	               &	               & 		             & 		             \\
122447.6-124852 & 20.12$\pm$0.06 & 20.32$\pm$0.06 & 20.05$\pm$0.07	& 19.93$\pm$0.06	& 19.81$\pm$0.07 & 19.57$\pm$0.06 & 19.77$\pm$0.06 \\
122448.2-122137 & 21.86$\pm$0.09 & 21.84$\pm$0.09 & 21.78$\pm$0.12	& 21.60$\pm$0.08	& 21.69$\pm$0.13 & 21.60$\pm$0.10 & 21.68$\pm$0.10 \\
122448.5-122847 & 19.67$\pm$0.06 & 19.64$\pm$0.06 & 19.48$\pm$0.06	& 19.39$\pm$0.06	& 19.37$\pm$0.06 & 19.26$\pm$0.06 & 19.33$\pm$0.06 \\
122448.8-124703 & 20.28$\pm$0.06 & 20.31$\pm$0.06 & 20.10$\pm$0.07	& 19.91$\pm$0.06	& 19.91$\pm$0.07 & 19.92$\pm$0.07 & 20.07$\pm$0.07 \\
\hline
\end{tabular}

%% file: tab7a.tex



\caption[]{\label{tab:rsh} Spectral types and redshifts of 173 objects 
           in the OACDF2}

\small

\begin{tabular}{llll}  \hline

OACDF           &           Type      &z$_{spec}$&   Comments          \\
                &                      &         &                     \\
                &                      &         &                     \\   \hline
                &                      &         &                     \\
122325.1-123757 &   early type ?       & 0.355 ? & S/N~3               \\
122325.4-123915 &   M2 star            &         & S/N~100             \\
122327.4-123714 &   early type         & 0.437   & S/N~5, emiss.       \\
122327.5-123718 &   early type         & 0.439   & S/N~15              \\
\hline

\end{tabular}

%% file: tab7b.tex
\normalsize

\begin{center}
 {{\bf Table 7.} Continued}
\end{center}

\vspace{0.3cm}

\small

\begin{tabular}{llll}  \hline

OACDF           &           Type      &z$_{spec}$& Comments             \\
                &                      &         &                      \\
                &                      &         &                      \\ \hline
                &                      &         &                      \\
122426.5-123002 &   G0-star            &         &  S/N~8               \\
122426.6-121947 &   A6-star            &         &  S/N~600             \\
122427.4-122313 &   M3-star            &         &  S/N~25              \\
122427.4-122924 &   G3-star            &         &  S/N~750             \\
\hline

\end{tabular}

%% file: tab7c.tex
\normalsize

\begin{center}
 {{\bf Table 7.} Continued}
\end{center}

\vspace{0.3cm}

\small

\begin{tabular}{llll}  \hline

OACDF           &           Type      &z$_{spec}$& Comments             \\
                &                      &         &                      \\
                &                      &         &                      \\   \hline
                &                      &         &                      \\
122447.6-124852 &   early type  ?      & 0.200 ? &  S/N~4               \\
122448.2-122137 &   ?                  & 0.434 ? &  S/N~3               \\
122448.5-122847 &   early type         & 0.316   &  S/N~15              \\
122448.8-124703 &   early type ?       & 0.159 ? &  S/N~4               \\
\hline

\end{tabular}